\documentclass[12pt]{iopart}

\usepackage{amssymb}		
\usepackage{braket}		
\usepackage{color}
\usepackage{graphicx}

\usepackage{units}

\usepackage{array}
\newcolumntype{x}[1]{>{\centering\arraybackslash\hspace{0pt}}p{#1}}

\usepackage{cite}

\newcommand{\binom}[2]{{#1 \choose #2}}
\newcommand{\eqref}[1]{(\ref{#1})}

\newcommand{\creation}[1]{b^{\dagger}_{#1}}
\newcommand{\annihilation}[1]{b^{\hphantom{\dagger}}_{#1}}

\newcommand{\JU}{J/U}
\newcommand{\Jcrit}{\big(\JU\big)_{\rm c}}
\newcommand{\Jcritapprox}[1]{\big(\JU\big)_{c,#1}}
\newcommand{\JcritQMC}{\Jcritapprox{\rm QMC}}

\newcommand{\muU}{\mu/U}
\newcommand{\numax}[1]{\nu_{\max,#1}}

\begin{document}

\title[Hypergeometric continuation, Shanks transformation, and Pad\'e 
       approximation]
      {Hypergeometric continuation of divergent perturbation series. \\
      II. Comparison with Shanks transformation and Pad\'e approximation}

\author{S\"oren Sanders$^1$, Martin Holthaus$^1$}

\address{$^1$ Institut f\"ur Physik, Carl von Ossietzky Universit\"at, D-26111 Oldenburg, Germany}

\ead{soeren.sanders@uni-oldenburg.de}

\date{\today}

\begin{abstract}
We explore in detail how analytic continuation of divergent perturbation series by generalized hypergeometric functions is achieved in practice. Using the example of strong-coupling perturbation series provided by the two-dimensional Bose-Hubbard model, we compare hypergeometric continuation to Shanks and Pad\'e techniques, and demonstrate that the former yields a powerful, efficient and reliable alternative for computing the phase diagram of the Mott insulator-to-superfluid transition. In contrast to Shanks transformations and Pad\'e approximations, hypergeometric continuation also allows us to determine the exponents which characterize the divergence of correlation functions at the transition points. Therefore, hypergeometric continuation constitutes a promising tool for the study of quantum phase transitions.
\end{abstract}




\maketitle 


\section{Introduction}
\label{sec:1}

Quantum mechanical perturbation theory usually is applied to a Hamiltonian of the form
\begin{equation}
 \widehat{H} = \widehat{H}_0 + \lambda \widehat{V} \; ,
\end{equation}
where the ``unperturbed system'' $\widehat{H}_0$ can be diagonalized exactly, and the ``perturbation'' $\widehat{V}$ is supposed to be small in some suitable sense~\cite{Kato95}. The dimensionless parameter~$\lambda$, which is set equal to one at the end of the calculation, connects the perturbed system~$\widehat{H}$ to the unperturbed one when varying between zero and unity. Employing the customary Rayleigh-Schr\"odinger perturbation series, one may then compute, e.g., the perturbed energy eigenvalues and eigenstates as a formal power series in $\lambda$~\cite{Rayleigh94,Schroedinger26,LaLiIII}. In practice, the evaluation of the perturbation series may pose insurmountable technical difficulties in higher orders, so that one often restricts oneself to the lowest few terms, hoping that the series converges sufficiently fast for such a truncation to be meaningful.\\
If, however, the perturbation series has only a finite radius of convergence, the formal series may still bear significance even beyond that radius. As an illustrative example, consider the one-dimensional linear harmonic oscillator~\cite{LaLiIII}  
\begin{equation}
 \widehat{H}_0 = \frac{\widehat{p}^2}{2m} + \frac{1}{2}m\omega_0^2 {\widehat{q}}^2 \; ,
\end{equation}
where~$\widehat{q}$ is the position operator, $\widehat{p}$ its conjugate momentum operator, $m$ denotes the mass of the oscillator particle, and $\omega_0$ is the (positive) oscillator angular frequency, so that the unperturbed energy eigenvalues are given by
\begin{equation}
 E_n(\lambda = 0) \; = \; \hbar\omega_0(n + 1/2)
\end{equation}
with quantum numbers $n = 0$, $1$, $2$, \ldots~. Suppose further that this system is perturbed by another oscillator potential with (positive) frequency 
$\omega_1$,   
\begin{equation}
 \widehat{V} = \frac{1}{2} m\omega_1^2 \widehat{q}^2 \; .
\end{equation}
Evidently, the {\em exact\/} perturbed energy eigenvalues then are given by 
\begin{equation}
 E_n(\lambda = 1) \; = \; \hbar\Omega(n + 1/2)
\label{eq:EXR}
\end{equation}
with
\begin{equation}
 \Omega = \sqrt{\omega_0^2 + \omega_1^2} \; .
\end{equation}	
If, on the other hand, one works out the perturbation series to fourth order, one finds
\begin{equation}
 E_n(\lambda = 1) \; = \; \hbar\omega_0(n + 1/2) \; f\big( (\omega_1/\omega_0)^2 \big) \; ,
\end{equation}
where the function~$f$ is given by
\begin{equation}
 f(x) = 1 + \frac{1}{2}x - \frac{1}{8}x^2 + \frac{1}{16} x^3 - \frac{5}{128} x^4 + {\mathcal O}(x^5) \; .
\end{equation}
Now one faces two closely related tasks. Firstly, one needs to deduce that this function~$f$, completed to all orders of~$x$, takes the form
\begin{equation}\label{eq:FIN}
 f(x) = \sum_{\nu=0}^\infty \binom{1/2}{\nu} \; x^\nu \; .
\end{equation}
This series converges for $|x| \leq 1$, but becomes a diverging, asymptotic series for $|x| > 1$~\cite{Hardy49,Copson65}. Therefore, the full perturbation 
series  
\begin{equation}
 E_n(\lambda = 1) \;
  = \; \hbar\omega_0(n + 1/2) \sum_{\nu=0}^\infty \binom{1/2}{\nu} \left( \frac{\omega_1}{\omega_0} \right)^{2\nu} 
\end{equation}
converges only for $\omega_1 \leq \omega_0$. Secondly, bearing in mind that
\begin{equation}\label{eq:ANC}
 \sum_{\nu=0}^\infty \binom{1/2}{\nu} x^\nu = \sqrt{1 + x} \qquad {\rm for} \quad |x| \leq 1 \; ,
\end{equation}
one has to realize that the expression on the right-hand side of this equation also constitutes the analytic continuation of the left-hand side for $x > 1$. Therefore, for both $\omega_1 \leq \omega_0$ and $\omega_1 > \omega_0$ the summed perturbation series yields
\begin{equation}
 E_n(\lambda = 1) \; = \; \hbar\omega_0(n + 1/2) \sqrt{1 + \left(\frac{\omega_1}{\omega_0}\right)^2} \; ,
\end{equation}
which, of course, equals the above expression~(\ref{eq:EXR}).\\
This pedagogical example sets the stage for the current work. The plan of the present paper is to subject a recently suggested powerful technique for the analytic continuation of divergent perturbation series based on the use of generalized hypergeometric functions~\cite{MeraEtAl15,PedersenEtAl16a,SandersEtAl15} --- dubbed {\em hypergeometric continuation\/} for short --- to a comprehensive test, thereby demonstrating its outstanding value for practical calculations.\\
Instead of aiming for general theorems, here we consider a definite system of particular significance, the two-dimensional Bose-Hubbard model. This model, which describes interacting Bose particles on an infinite two-dimensional lattice, shows a quantum phase transition from a Mott insulator to a superfluid at zero temperature when the relative strength of the interparticle interaction is reduced, or the strength of the tunneling contact between neighboring lattice sites is increased~\cite{FisherEtAl89,Sachdev99}. This Mott insulator-to-superfluid transition reflects itself in a divergence of the strong-coupling perturbation series of certain correlation functions of the Bose-Hubbard model~\cite{SandersEtAl15}. Therefore, in order to obtain information on system properties in the superfluid phase, these series need to be analytically continued beyond the point of divergence into the superfluid regime. Hence, we again face the two tasks exemplified by equations~(\ref{eq:FIN}) and (\ref{eq:ANC}): Starting from a series of which only the first few terms are at our disposal, we need to guess the underlying systematics to all orders, and to deduce the corresponding analytic continuation. In the Bose-Hubbard case, however, the leading contributions to the perturbation series are given only numerically, and their extension to infinite orders is not obvious.\\
For tackling these tasks, we proceed as follows: In Sec.~\ref{sec:2} we briefly review the formulation of the Bose-Hubbard model, and give a formal definition of its $k$-particle correlation functions $c_{2k}$ in terms of strong-coupling perturbation series. In Sec.~\ref{sec:3} we then explain the basics of their analytic continuation by means of the familiar Shanks transformation and Pad\'e approximation methods~\cite{Shanks55,BenderOrszag99,BakerMorris10,CalicetiEtAl07}, and through the novel hypergeometric technique. All three schemes are applied in Sec.~\ref{sec:4} to the one-particle correlation function $c_2$ of the Bose-Hubbard model, in order to deduce its phase diagram. Since there exist fairly precise previous computations of this diagram, comparison of such reference data with our results allows us to gauge the accuracy of the respective method, and thereby to confirm that hypergeometric continuation is at least competitive with its well-established rivals. In Sec.~\ref{sec:5} we then address a subject which highlights a particular strength of hypergeometric continuation, and which is {\em not\/} amenable to any of the other two methods: Using hypergeometric continuation, we determine the exponents with which the one-particle correlation function $c_2$ and the two-particle correlation function $c_4$ diverge at the phase boundary. Finally, some conclusions are drawn in Sec.~\ref{sec:6}.\\
Thus, the present case study serves a twofold purpose. On a fairly general level, we wish to establish hypergeometric continuation as a reliable and highly versatile tool for extracting observable quantities from divergent perturbation series, corroborating the pioneering work by Mera \etal~\cite{MeraEtAl15,PedersenEtAl16a}. More specifically, the results obtained in Sec.~\ref{sec:5} also serve as input data for our preceding paper~\cite{SandersHolthaus17a}, in which we have determined the critical exponent of the order parameter of the Mott transition on a two-dimensional lattice. While the present investigation is essentially independent of that preceding paper~\cite{SandersHolthaus17a}, it provides the technical background information required for assuring the correctness of the data employed therein.
\section{Perturbative evaluation of correlation functions}
\label{sec:2}
\begin{figure}[t]
 \begin{center}
  \includegraphics[width = 0.8\textwidth]{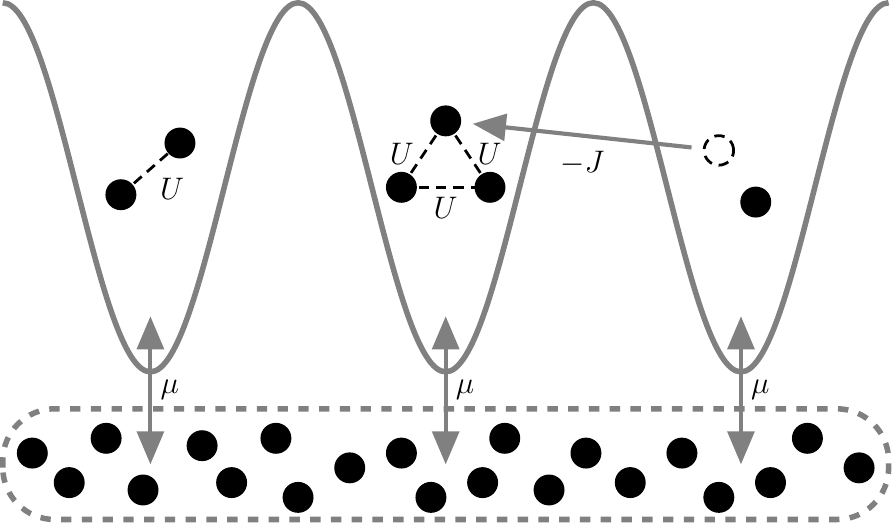}
 \end{center}
 \caption{Schematic visualization of the Bose-Hubbard model with on-site repulsion energy~$U$, hopping matrix element~$J$, and chemical potential~$\mu$.}      
 \label{F_1}
\end{figure}

\begin{figure}[t]
 \begin{center}
  \includegraphics[width = 0.8\textwidth]{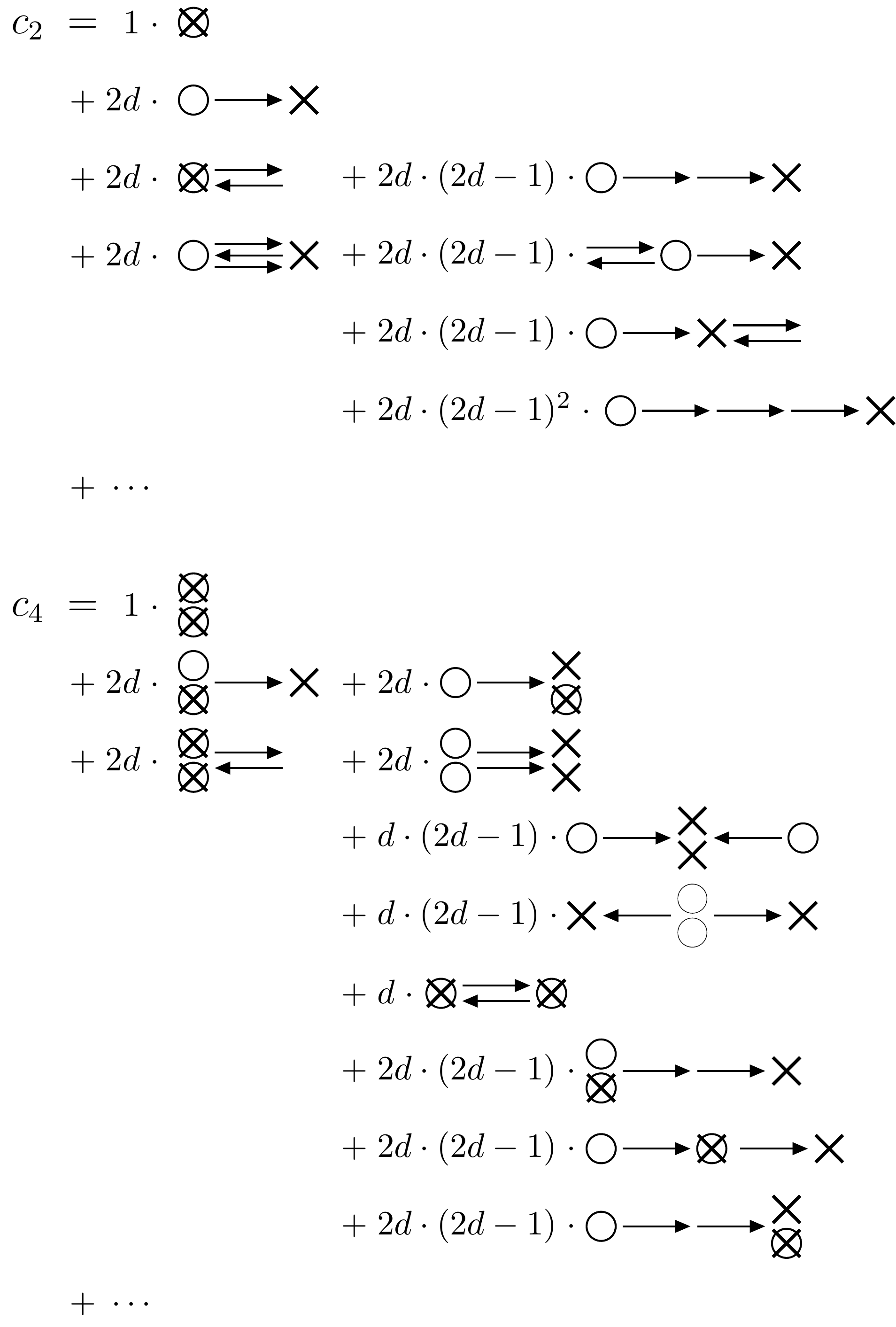}
 \end{center}
 \caption{Graphical representation of the lowest terms of the perturbation series~\eqref{eq:series_expansion_of_c_2k} for the one-particle correlation function~$c_2$ (above), and for the two-particle correlation function~$c_4$ (below), with weight factors pertaining to a $d$-dimensional hypercubic lattice. Open circles symbolize a creation process, crosses an annihilation process, and arrows indicate a tunneling event which connects neighboring lattice sites.}
 \label{F_2}
\end{figure}

\begin{figure}[t]
\begin{center}
\includegraphics[width = 0.8\textwidth]{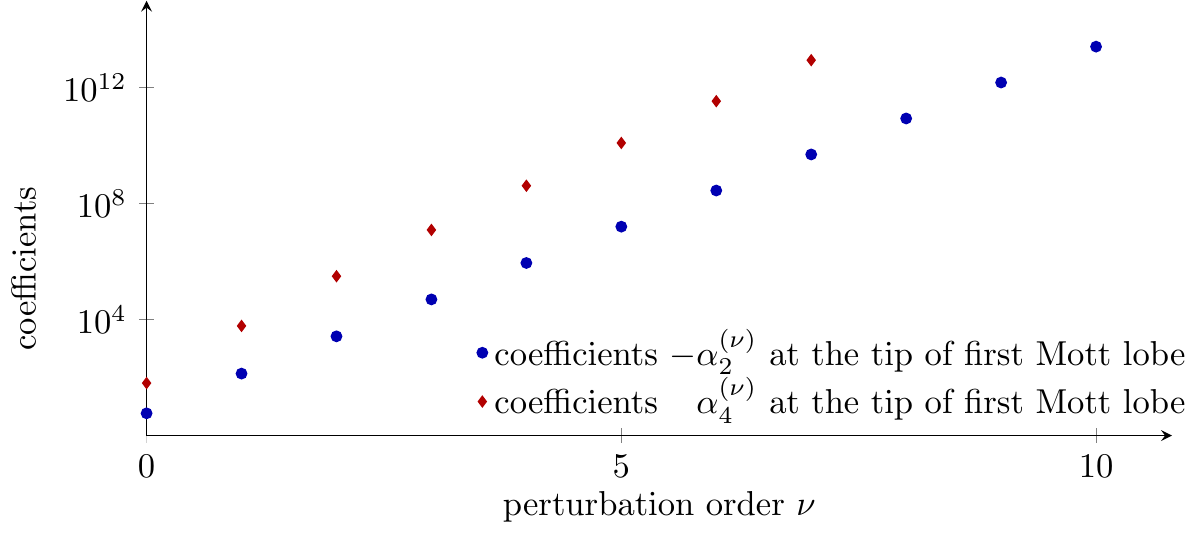}
\end{center}
\caption{(Color online) Coefficients $-\alpha_2^{(\nu)}(\mu/U)$ and $\alpha_4^{(\nu)}(\mu/U)$ of the series~\eqref{eq:series_expansion_of_c_2k} for $\mu/U = 0.3769$, as corresponding to a chemical potential near the tip of the lowest Mott lobe for $d = 2$ shown in Figs.~\ref{fig:phase_diagram_comparison_with_Shanks_2d},\ref{fig:phase_diagram_comparison_with_Pade_2d},\ref{fig:phase_diagram_2d}. Observe the logarithmic scale of the ordinate!}
\label{F_3}
\end{figure}

Consider a $d$-dimensional lattice of arbitrary geometry with sites labeled by an index~$i$, and let $\creation{i}$ and $\annihilation{i}$ be the Fock space operators which create and annihilate, respectively, a Bose particle at the $i$th site, obeying the commutation relation  $[\annihilation{i},\creation{j}]= \delta_{ij}$, so that  $\widehat{n}_i = \creation{i}\annihilation{i}$ is the operator which counts the number of particles placed on site~$i$. Assume further that neighboring sites are connected by a tunneling contact with hopping matrix element $J$, and that particles occupying a common site repel each other, with each pair of particles contributing an amount~$U$ to the total potential energy, while interaction between particles sitting on different sites is neglected, as sketched in Fig.~\ref{F_1}. This system is subjected to a chemical potential~$\mu$, which enables one to control its total particle content. After scaling with respect to~$U$, the dimensionless Hamiltonian of this Bose-Hubbard model is cast into the form
\begin{equation}\label{eq:BaseH}
 \widehat{H}_{\rm BH} =  \widehat{H}_0 + \widehat{H}_{\rm tun} \; ,
\end{equation}
where the first part
\begin{equation}\label{eq:SiteD}
 \widehat{H}_0 = \frac{1}{2} \sum_{i} \widehat{n}_i(\widehat{n}_i-1) - \mu/U \sum_{i} \widehat{n}_i
\end{equation}
is site-diagonal, accounting for the total repulsion energy and the coupling to the chemical potential, while the second term
\begin{equation}\label{eq:TunnL}
 \widehat{H}_{\rm tun} = -J/U \sum_{\langle i,j \rangle} \, \creation{i} \annihilation{j}
\end{equation}
incorporates the tunneling links between neighboring sites~$i$ and $j$, as indicated by the symbol~$\langle i,j \rangle$. This fairly minimalistic model provides an excellent description of ultracold atoms in deep optical lattice potentials~\cite{JakschEtAl98,GreinerEtAl02}, since the effective range of the two-particle scattering potential for neutral atoms is much shorter than the optical lattice constant, as given by half the wavelength of the lattice-generating laser radiation~\cite{BlochEtAl08}.\\
Assuming integer filling with $g$ particles per site at zero temperature, the ground state of the system in the absence of the tunneling contact, that is, for $J/U = 0$, is given by the product state
\begin{equation}\label{eq:MottS}
 \ket{g} = \prod_i \frac{(\creation{i})^g}{\sqrt{g!}} \ket{\rm vac}
\end{equation}
with $\mu/U < g < \mu/U + 1$~\cite{FisherEtAl89}, and $\ket{\rm vac}$ denoting the empty-lattice state. Now there are two alternative approaches to account for finite tunneling strength by means of perturbation theory in $J/U$. The first strategy is to compare the energy of the ground state which emerges from the product state~\eqref{eq:MottS} with the energy of two defect states which contain either an additional particle or an additional hole moving coherently over the lattice; the boundary between the incompressible Mott phase and the superfluid phase is reached when the energy difference between the Mott state and a defect state vanishes. This strategy has been adopted and implemented to third order already in Ref.~\cite{FreericksMonien94}. The technical inconvenience of having to deal with the defect states is avoided if instead one probes the response of the basic system~\eqref{eq:BaseH} to external sources and drains, thus breaking its particle number conservation. Taking these sources and drains to be spatially uniform with strength~$\eta$, we therefore consider the extended system
\begin{equation}\label{eq:ExtnS}
 \widehat{H} = \widehat{H}_{\rm BH} + \widehat{H}_{\rm sd} \; ,
\end{equation}  
where     
\begin{equation}\label{eq:HsrcD}
 \widehat{H}_{\rm sd} = \sum_i \eta \left( \creation{i} + \annihilation{i} \right) \; , 
\end{equation}
and determine its intensive ground state energy ${\mathcal E}$ in the thermodynamic limit, 
\begin{equation}\label{eq:IntGE}
 {\mathcal E}(\mu/U, J/U, \eta) = \lim_{M \to \infty} \braket{\widehat{H}} / M \; ,
\end{equation}
where $M$ is the number of lattice sites. Observing that the sought-for response then is quantified by 
\begin{equation}
 \frac{\partial {\mathcal E}}{\partial\eta} = 2 \braket{\annihilation{i}}
\end{equation}
for arbitrary~$i$, where the expectation value again is taken with respect to the ground state of the extended system~(\ref{eq:ExtnS}), the responseless Mott phase is characterized by $\braket{\annihilation{i}} = 0$, whereas one finds $\braket{\annihilation{i}} \neq 0$ in the superfluid phase, reflecting spontaneous symmetry breaking.\\
Expanding the ground state energy~(\ref{eq:IntGE}) within the Mott phase, which has to be an even function of $\eta$, in the form
\begin{equation}
 {\mathcal E}(\mu/U, J/U, \eta) = e_0(\mu/U , J/U) + \sum_{k=1}^\infty c_{2k}(\mu/U, J/U) \, \eta^{2k} \; ,
\end{equation}
the quantity $e_0(\mu/U , J/U)$ merely represents the intensive ground state energy of the basic model~(\ref{eq:BaseH}), whereas the desired information is contained in the $k$-particle correlation functions~$c_{2k}(\mu/U, J/U)$. When these correlation functions, in their turn, are expanded for given $\mu/U$ in powers of the scaled hopping strength $J/U$, writing
\begin{equation}\label{eq:series_expansion_of_c_2k}
 c_{2k}(\mu/U, J/U) = \sum_{\nu = 0}^\infty \alpha_{2k}^{(\nu)}(\mu/U) \, \big(J/U\big)^\nu \; ,
\end{equation}
each term in such a series corresponds to various chains of $k$~creation processes, $\nu$~tunneling events, and $k$ annihilation processes, as visualized in Fig.~\ref{F_2} for $k = 1$ and $k = 2$, respectively. We pay particular attention to these two correlation functions $c_2$ and $c_4$: When varying $\JU$ at fixed $\muU$ close to the respective transition point $\Jcrit$, they exhibit power-law behaviors
\begin{equation}\label{eq:introduction_of_the_divergence_exponent_c_2}
 c_2 \sim \Big(\Jcrit - \JU\Big)^{-\epsilon_2(\muU)} \qquad {\rm for} \quad \JU \rightarrow \Jcrit
\end{equation}
and
\begin{equation}\label{eq:introduction_of_the_divergence_exponent_c_4}
 c_4 \sim \Big(\Jcrit - \JU\Big)^{-\epsilon_4(\muU)} \qquad {\rm for} \quad \JU \rightarrow \Jcrit
\end{equation}
with positive divergence exponents $\epsilon_2(\muU)$ and $\epsilon_4(\muU)$ which enable one, for $d = 2$, to determine the critical exponent of the order parameter~\cite{SandersHolthaus17a}.\\
From the viewpoint of systematic many-body perturbation theory, this approach means that the Hamiltonian~(\ref{eq:ExtnS}) is split according to  
\begin{equation}
 \widehat{H} = \widehat{H}_0 + \widehat{V} \; ,
\end{equation}
where the ``unperturbed system''~$\widehat{H}_0$ is given by the site-diagonal part~(\ref{eq:SiteD}), and the ``perturbation''
\begin{equation}
 \widehat{V} = \widehat{H}_{\rm tun} + \widehat{H}_{\rm sd}
\end{equation}
consists of both the tunneling contacts~\eqref{eq:TunnL} and the sources and drains~\eqref{eq:HsrcD}. The perturbative evaluation of the series~\eqref{eq:series_expansion_of_c_2k}, starting from a ground state~\eqref{eq:MottS} with appropriate filling factor~$g$ for $J/U = \eta = 0$, then proceeds by means of the process chain approach as devised by Eckardt~\cite{Eckardt09}, which constitutes an adaption of Kato's non-recursive formulation of the general perturbation series~\cite{Kato49} to many-body lattice models. This enables one to represent physical observables by appropriate sequences of processes as exemplified above in Fig.~\ref{F_2}; the evaluation of such diagrams is sketched briefly in \ref{appendix:analytic_calculations_in_low_order}. The $n$th order of perturbation theory thus comprises all connected diagrams which consist of $n$~processes of any kind; the computational bottleneck is caused by the necessity to evaluate all processes of a given chain in all possible permutations. The details of this procedure have been communicated elsewhere~\cite{TeichmannEtAl09} and do not need to concern us here; suffice it to state that we are able to compute the series~\eqref{eq:series_expansion_of_c_2k} up to $\nu_{\max,2} = 10$ for $k = 1$ and up to $\nu_{\max,4} = 7$ for $k = 2$ (corresponding, respectively, to orders $n = 12$ and $n = 11$ of perturbation theory). Thus, in the present study we fully account for all process chains contributing to the one-particle correlation function $c_2$ with up to ten tunneling events, and for all chains contributing to $c_4$ with up to seven tunneling events.\\
As an example, Fig.~\ref{F_3} shows the coefficients $-\alpha_2^{(\nu)}(\mu/U)$ and $\alpha_4^{(\nu)}(\mu/U)$ for $\mu/U = 0.3769$, as corresponding to a scaled chemical potential near the tip of the lowest Mott lobe showing up in the phase diagram (see the later Figs.~\ref{fig:phase_diagram_comparison_with_Shanks_2d},\ref{fig:phase_diagram_comparison_with_Pade_2d},\ref{fig:phase_diagram_2d}). Observing the logarithmic scale of the ordinate, one deduces from the exponential growth of the coefficients that simply terminating the series~\eqref{eq:series_expansion_of_c_2k} at the respective $\nu_{\max,2k}$ will not yield reasonable approximations. Hence, we need some kind of extrapolation scheme for estimating the coefficients $\alpha_{2k}^{(\nu)}(\mu/U)$ for $\nu > \nu_{\max,2k}$, and some means of analytic continuation for giving meaning to the series beyond their radius of convergence.
\section{Analytic continuation}
\label{sec:3}
We start by inspecting two standard techniques for handling slowly converging or diverging series: the Shanks transformation and the Pad\'e approximation, before turning to analytic continuation utilizing hypergeometric functions. We discuss the three methods using $c_2$ as example; $c_4$ can be treated analogously.
\subsection{Shanks transformation}
\begin{figure}[tb]
 \begin{center}
  \includegraphics[width=0.8\textwidth]{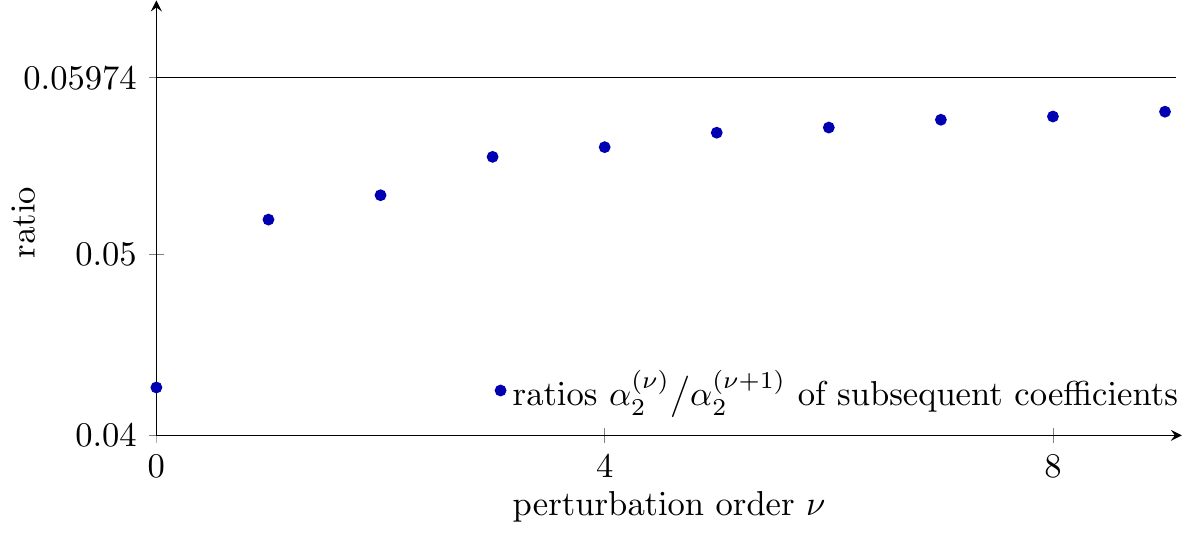}
 \end{center}
 \caption{(Color online) Ratios $\Jcritapprox{\nu} = {\alpha_2^{(\nu)}}\big/{\alpha_2^{(\nu+1)}}$ of subsequent coefficients of the series~\eqref{eq:series_expansion_of_c_2k} as approximants for the phase boundary for $d = 2$ and $\muU = 0.3769$, as corresponding to a chemical potential near the tip of the lowest Mott lobe shown in Figs.~\ref{fig:phase_diagram_comparison_with_Shanks_2d},\ref{fig:phase_diagram_comparison_with_Pade_2d},\ref{fig:phase_diagram_2d}. The solid horizontal line indicates the limit $\JcritQMC = 0.05974(3)$ expected from QMC calculations~\cite{CapogrossoSansoneEtAl08}.}
 \label{F_4}
\end{figure}

\begin{figure}[tb]
 \begin{center}
  \includegraphics[width=0.8\textwidth]{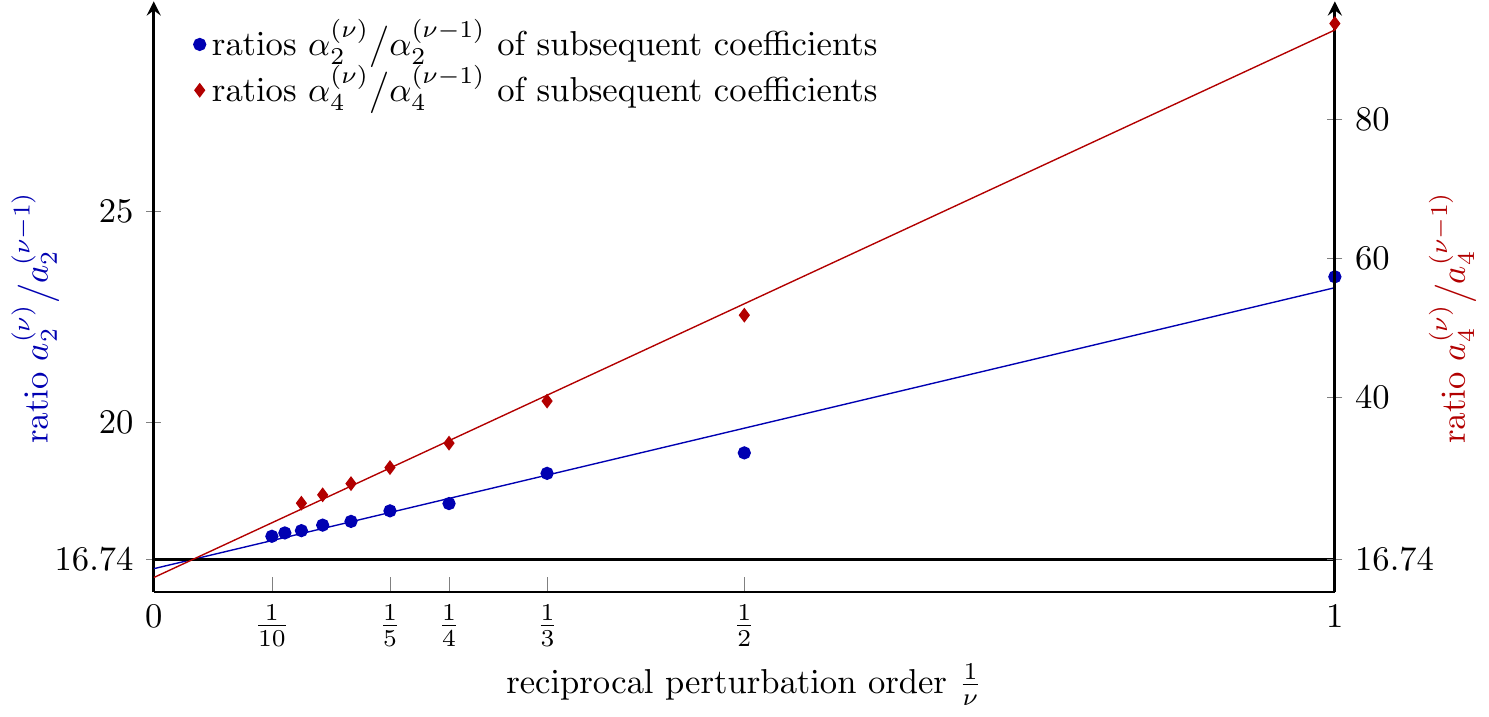}
 \end{center}
  \caption{(Color online) Ratios ${\alpha_2^{(\nu)}}\big/{\alpha_2^{(\nu-1)}}$ and ${\alpha_4^{(\nu)}}\big/{\alpha_4^{(\nu-1)}}$ of subsequent coefficients of the series~\eqref{eq:series_expansion_of_c_2k} for $d = 2$ and $\mu/U = 0.3769$, plotted vs.\ the reciprocal perturbation order $1/\nu$, together with linear fits in the reciprocal order $1/\nu$. Again, the horizontal line indicates the limit $1\big/\JcritQMC = 1\big/0.05974(3)$ expected from QMC calculations~\cite{CapogrossoSansoneEtAl08}. Observe that the limits $16.51$ and $14.07$ for $1/\nu \rightarrow 0$ of the linear fits to the ratios ${\alpha_2^{(\nu)}}\big/{\alpha_2^{(\nu-1)}}$ and ${\alpha_4^{(\nu)}}\big/{\alpha_4^{(\nu-1)}}$, respectively, differ notably.}
 \label{F_5}
\end{figure}

We first consider the Shanks transformation~\cite{Shanks55,BenderOrszag99}, which transforms a sequence $(s_n)_{n\in\mathbb{N}}$ into the sequence $(S_n)_{n\in\mathbb{N}}$ given by
\begin{equation}\label{eq:definition_of_the_Shanks_transformation}
 S_n := \frac{s_{n+1} \cdot s_{n-1} -s_n^2}{s_{n+1} - 2 s_n + s_{n-1}} \; .
\end{equation}
The Shanks transformation is constructed in such a way that for $s_n = A + \alpha q^n$ with geometrically decaying transient ($|q| < 1$) the Shanks transform $(S_n)_{n\in\mathbb{N}}$ is a constant sequence $S_n = A$ which equals the limit of the original sequence $(s_n)_{n\in\mathbb{N}}$.\\
More generally, in many cases the Shanks transform has better convergence properties than the original sequence. Applied to the sequence of partial sums $s_n = \sum_{\nu = 0}^n b_\nu$ of some sequence $(b_\nu)_{\nu\in\mathbb{N}}$, the transformation formula~\eqref{eq:definition_of_the_Shanks_transformation} takes the form
\begin{equation}
 S_n = s_{n-1} - \frac{b_n^2}{b_{n+1} - b_n} \; .
\end{equation}
We now apply this general prescription to the truncation $s_n = \sum_{\nu = 0}^n \alpha_2^{(\nu)}(\muU) \, \big(\JU\big)^\nu$ of the perturbation series for $c_2$. Since we always monitor the phase transition such that $\muU$ is kept fixed while $\JU$ is varied, we omit the argument $\muU$ in the following for the sake of clear notation. Thus we obtain
\begin{equation}\label{eq:Shanks_transform_of_c2}
 \eqalign{
   S_n
  &= s_{n-1} - \frac{\left( \alpha_2^{(n)} \, \big(\JU\big)^n\right)^2}{\alpha_2^{(n+1)}\, \big(\JU\big)^{n+1} - \alpha_2^{(n)}\, \big(\JU\big)^n} \\
  &= s_{n-1} - \frac{\big(\alpha_2^{(n)}\big)^2 \, \big(\JU\big)^n}{\alpha_2^{(n+1)} \cdot \JU - \alpha_2^{(n)}} \; .}
\end{equation}
The underlying hypothesis now is that the limit of this Shanks transform~\eqref{eq:Shanks_transform_of_c2} is $c_2$, the object we are interested in. Then approximants to the radius of convergence of $c_2$ are given by the respective zero of the denominator of the second term, so that we have the easy-to-calculate approximation
\begin{equation}\label{eq:Shanks_transform_phase_boundary}
 \Jcritapprox{\nu} = \frac{\alpha_2^{(\nu)}}{\alpha_2^{(\nu+1)}}
\end{equation}
for the phase boundary $\Jcrit = \lim_{\nu\rightarrow\infty} \Jcritapprox{\nu}$.
Exemplarily we have displayed these ratios for the tip of the lowest Mott lobe in Fig.~\ref{F_4}. The approximants $\Jcritapprox{\nu}$ seem to converge towards a limit consistent with the corresponding phase boundary $\JcritQMC$ obtained from Quantum Monte Carlo (QMC) calculations~\cite{CapogrossoSansoneEtAl08}.
Since the limit is not reached in finite order, we have linearly extrapolated the inverse of these finite-order approximants over the reciprocal perturbation order $1/\nu$ to the limit $1/\nu \rightarrow 0$, as displayed in Fig.~\ref{F_5}. This latter figure also shows the ratios ${\alpha_4^{(\nu)}}\big/{\alpha_4^{(\nu-1)}}$ corresponding to the two-particle correlation function $c_4$. Observe that the fit then yields a notably different numerical value for $\Jcrit$, although we expect the same $\Jcrit$ for both $c_2$ and $c_4$~\cite{SandersHolthaus17a}. This indicates that the present method is not optimal.\\
Although the Shanks transformation thus allows us to compute the phase boundary, it does not provide access to the divergence exponents $\epsilon_{2k}$, as introduced in the relations~\eqref{eq:introduction_of_the_divergence_exponent_c_2} and \eqref{eq:introduction_of_the_divergence_exponent_c_4}, because the transform~\eqref{eq:Shanks_transform_of_c2} has a pole of order~$1$. Therefore, the Shanks transform yields only trivial, integer-valued estimates of the divergence exponents $\epsilon_{2k}$, independent of the perturbation coefficients $\alpha_{2k}^{(\nu)}(\muU)$.\\
We remark that in some cases it is possible to further increase the convergence properties of the Shanks transform $S_n$ by iterating the Shanks transformation. While this may lead to a better approximation of $c_2$ in general, its radius of convergence and consequently our estimate $\Jcritapprox{\nu}$ for the phase boundary are not affected by such an iteration. This is a consequence of the fact that any singularity of $s_n$ is as well a singularity of the Shanks transform $S_n$, due to the numerator's quadratic and the denominator's linear dependence on $s_n$, so that the Shanks transform of $S_n$ shares the singularity of $S_n$ itself at $\Jcritapprox{\nu} = \alpha_2^{(\nu)}\big/\alpha_2^{(\nu+1)}$.\\
To support this theoretical discussion, we have sketched in \ref{appendix:analytic_calculations_in_low_order} the analytical calculation of the first three coefficients $\alpha_2^{(0)}$, $\alpha_2^{(1)}$, and $\alpha_2^{(2)}$ of $c_2$, in order to evaluate $S_1$. In the limit of high dimensionality the previously derived formula~\eqref{eq:Shanks_transform_phase_boundary} then reproduces the mean-field phase boundary, which is exact in this limit, thus serving as a showcase for the quality of the Shanks transformation.
\subsection{Pad\'e approximation}
Next, we address the Pad\'e approximation~\cite{BenderOrszag99,BakerMorris10,CalicetiEtAl07}. The underlying hypothesis here is that the coefficients $\alpha_{2k}^{(\nu)}$ of the perturbation series correspond to coefficients of the series expansion of a rational function
\begin{equation}\label{eq:definition_of_the_Pade_approximation}
 A_{L/M}(\JU) = \frac{\sum_{k=0}^L p_k \cdot \big(\JU\big)^k}{1 + \sum_{k=1}^M q_k \cdot \big(\JU\big)^k} \; .
\end{equation}
Given the degrees $L$ and $M$ of the polynomials in the denominator and numerator of $A_{L/M}$, the coefficients $p_0,p_1,...,p_L$ and $q_1,...,q_M$ of the polynomials can be calculated by solving the system of equations
\begin{equation}
 \eqalign{
  \alpha_{2k}^{(0)} &= A_{L/M}(0) \hspace{42pt} = p_0\\
  \alpha_{2k}^{(1)} &= \frac{\partial A_{L/M}}{\partial (\JU)}\Bigg|_{J/U=0} \hspace{17pt} = p_1 - p_0 q_1\\
  \alpha_{2k}^{(2)} &= \frac{1}{2} \frac{\partial^2 A_{L/M}}{\partial (\JU)^2}\Bigg|_{J/U=0} = p_2 - p_0 q_2 - (p_1 - p_0 q_1) q_1\\
                 & \hspace{7pt} \vdots
 }
\end{equation}
The degrees $L$ and $M$ are restricted through the set of available coefficients $\alpha_{2k}^{(\nu)}$,
\begin{equation}
 (L + 1) + M \le 1 + \numax{2k} \; .
\end{equation}
To obtain an approximation to the phase boundary, as given by the radius of convergence of $c_2$, we observe that the rational function $A_{L/M}$ has up to $M$ isolated poles of integer order, corresponding to the zeros of the denominator $1 + \sum_{k=1}^M q_k \cdot \big(\JU\big)^k$. Thus, if we assume that $c_2$ is given by the Pad\'e approximation $A_{L/M}$, the smallest positive pole of $A_{L/M}$ corresponds to the phase boundary. Again, the integer order of the poles of $A_{L/M}$ implies that the Pad\'e approach only yields integer estimates for the divergence exponents.\\
Shanks transformation and Pad\'e approximation are directly related to each other. It can be shown that the Shanks transform $S_n$ corresponds to the Pad\'e approximation $A_{n/1}$, while Pad\'e approximations $A_{L/M}$ with $M \le L$ equal so-called generalized Shanks transforms~\cite{BenderOrszag99}.
\subsection{Hypergeometric continuation}
\begin{figure}[tb]
 \begin{center}
  \includegraphics[width=0.8\textwidth]{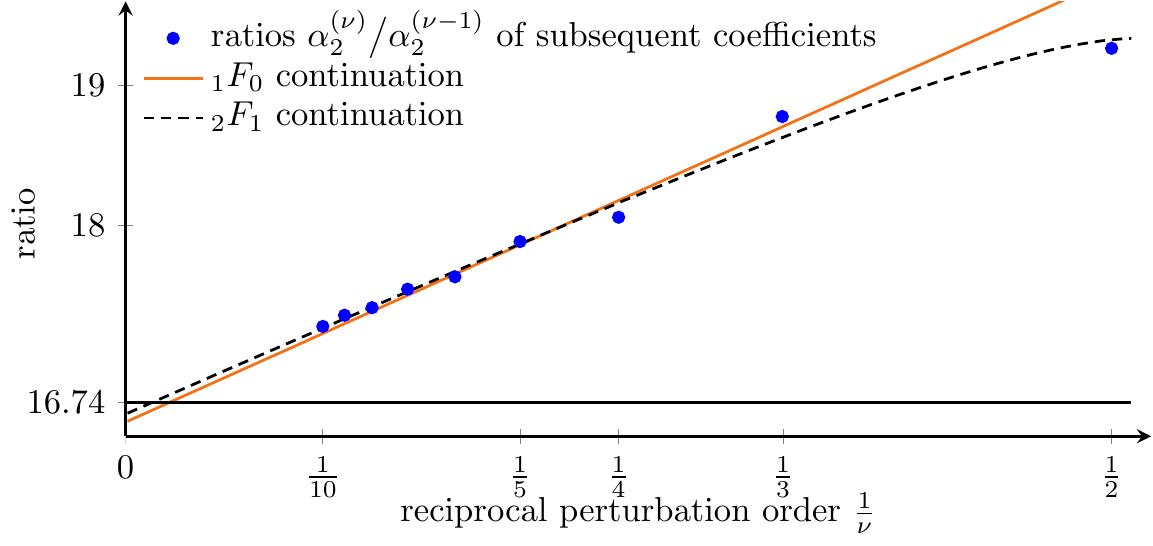}
 \end{center}
 \caption{(Color online) Ratios ${\alpha_2^{(\nu)}}\big/{\alpha_2^{(\nu-1)}}$ of subsequent coefficients of the series~\eqref{eq:series_expansion_of_c_2k} for $d = 2$ and $\mu/U = 0.3769$, plotted vs.\ the reciprocal order $1/\nu$, together with the corresponding fit to the binomial series, and to the Gaussian hypergeometric function. Again, the horizontal line indicates the limit $1\big/\JcritQMC = 1\big/0.05974(3)$ expected from QMC calculations~\cite{CapogrossoSansoneEtAl08}.}
 \label{F_6}
\end{figure}

Summing up, we stress that neither Shanks transformation nor Pad\'e approximation are able to yield non-trivial estimates for the divergence exponents~$\epsilon_{2k}$. The idea to overcome this deficiency is to replace the rational functions used in the Pad\'e approximation by functions with an essential singularity. In a recent series of papers~\cite{MeraEtAl15,PedersenEtAl16a}, Mera~\etal have suggested to employ hypergeometric functions for the analytic continuation of typical perturbation series in quantum mechanics. As we will discuss below, these hypergeometric functions possess a tunable singularity, which is exactly what is needed for our purposes. Moreover, we have already given a heuristic argument why hypergeometric functions are particularly well suited for the analytic continuation of strong coupling perturbation series for the Bose-Hubbard model~\cite{SandersEtAl15}.\\
For instance, we may assume that the coefficients $c_{2k}$ are given by Gaussian hypergeometric functions with parameters $a$, $b$, $c$ and $\Jcrit$ still to be determined,
\begin{equation}
 c_{2k} = \alpha_{2k}^{(0)} \cdot {_2F_1}\left(a,b;c;\frac{\JU}{\Jcrit}\right) = \alpha_{2k}^{(0)} \sum_{\nu = 0}^\infty \frac{(a)_\nu \, (b)_\nu}{\nu! \, (c)_\nu} \left(\frac{\JU}{\Jcrit}\right)^\nu \; ,
\end{equation}
where $(a)_\nu = a (a + 1) \cdots (a + \nu -1)$ is the usual Pochhammer symbol~\cite{AbramowitzStegun72}. 
Beyond this Gaussian hypergeometric function ${_2F_1}$ with $4$ degrees of freedom, we also consider generalized hypergeometric functions
\begin{equation}\label{eq:definition_of_the_generalized_hypergeometric_function}
 \eqalign{
 {_pF_q}\Bigg(a_1,a_2,...,a_p,b_1,...&,b_q,\frac{\JU}{\Jcrit}\Bigg)\\ &=\sum_{\nu=0}^\infty \frac{(a_1)_\nu \, (a_2)_\nu \cdots (a_p)_\nu}{\nu! \, (b_1)_\nu \, (b_2)_\nu \cdots (b_q)_\nu}
          \left(\frac{J/U}{(J/U)_{\rm c}}\right)^\nu \; .}
\end{equation}
It is known that such generalized hypergeometric functions converge for all arguments if $p < q +1$ and diverge for any value of $\JU \ne 0$ if $p > q + 1$~\cite{MagnusEtAl66}. As a finite radius of convergence is precisely what we are interested in, we employ only generalized hypergeometric functions with $p = q + 1$, as natural generalizations of ${_2F_1}$. As defined in equation~\eqref{eq:definition_of_the_generalized_hypergeometric_function}, these functions have $\Jcrit$ as their radius of convergence. In the following we will describe our approach for the Gaussian hypergeometric function ${_2F_1}$; the adjustments necessary for generalized hypergeometric functions ${_{q+1}F_q}$ are straightforward.\\
For calculating the parameters $a$, $b$, $c$ and $\Jcrit$ we utilize all perturbatively available coefficients $\alpha_{2k}^{(\nu)}$ in a least-squares fit as displayed in Fig.~\ref{F_6}. As the coefficients vary over at least thirteen orders of magnitude (cf.~Fig.~\ref{F_3}), an unweighted fit yields huge relative deviations for the smaller coefficients $\alpha_{2k}^{(\nu)}$, and has a correspondingly bad performance. To overcome this deficiency we have experimented with various weights. We obtained good results with least-squares fits of the relative deviations, thus weighting the absolute deviations by the respective coefficient $1/\alpha_{2k}^{(\nu)}$. However, we achieved the best overall performance when fitting the ratios $\alpha_{2k}^{(\nu)}\big/\alpha_{2k}^{(\nu-1)}$ of subsequent coefficients by corresponding ratios of terms of the hypergeometric function,
\begin{equation}\label{eq:ratio_in_the_ratio_fit}
 \frac{\frac{(a)_\nu\,(b)_\nu}{\nu!\,(c)_\nu}
          \left(\frac{1}{(J/U)_{\rm c}}\right)^\nu}{\frac{(a)_{\nu-1}\,(b)_{\nu-1}}{(\nu-1)!\,(c)_{\nu-1}}
          \left(\frac{1}{(J/U)_{\rm c}}\right)^{\nu-1}} = \frac{(a + \nu - 1)\,(b + \nu - 1)}{\nu \, (c + \nu - 1)} \cdot \frac{1}{(J/U)_{\rm c}} \; .
\end{equation}
These ratios are of the same order of magnitude, so that we do not have to introduce artificial weights when fitting the data. In \ref{appendix:comparison_between_ratio_and_weighted} we have compiled a comparison between both procedures, employing the ratios~\eqref{eq:ratio_in_the_ratio_fit} on the one hand, and relative deviations on the other, and show that both methods lead to more or less the same results.\\
Depending on the details of the fitting algorithm used, unwanted local minima of the sum of least squares can cause problems which worsen with increasing number of degrees of freedom. To deal with this we calculate every fit multiple times with randomized initial values, and choose the fit with the smallest squared deviations.
In Fig.~\ref{F_6} we have displayed the ratios $\alpha_2^{(\nu)}\big/\alpha_2^{(\nu-1)}$ of the coefficients near the tip of the first Mott lobe together with the fits we obtained for the Gaussian hypergeometric function ${_2F_1}$, and for the generalized hypergeometric function ${_1F_0}$, which is the well-known binomial series.\\
We now turn to the evaluation of the divergence exponents $\epsilon_2$ and $\epsilon_4$ of the correlation functions~\eqref{eq:introduction_of_the_divergence_exponent_c_2} and \eqref{eq:introduction_of_the_divergence_exponent_c_4}. To this end we apply the linear transformation formula (see, e.g.,~\cite{AbramowitzStegun72})
 \begin{eqnarray}\label{eq:linear_transformation_formula_for_hypergeometric_functions}
  \fl
  {_2F_1}(a,b;c;z) = \frac{\Gamma(c) \, \Gamma(c-a-b)}{\Gamma(c-a) \, \Gamma(c-b)} \; {_2F_1}(a,b;a+b-c+1;1-z)\nonumber \\
   + (1-z)^{c-a-b} \, \frac{\Gamma(c) \, \Gamma(a+b-c)}{\Gamma(a) \, \Gamma(b)} \; {_2F_1}(c-a,c-b;c-a-b+1;1-z)
 \end{eqnarray}
which is valid for any $z \in \mathbb{C}$ with $|\arg(1-z)| < \pi$. The hypergeometric functions ${_2F_1}$ on the right-hand side tend to $1$ as their arguments $1-z$ tend to $0$. Therefore, the first term is asymptotically constant for $z \rightarrow 1$. In contrast, for $a + b > c$ the factor $(1-z)^{c-a-b}$ diverges, and the asymptotics are given by
\begin{equation}
 \eqalign{
  {_2F_1}\left(a,b;c;\frac{\JU}{\Jcrit}\right)
   &\sim \left(1 - \frac{\JU}{\Jcrit}\right)^{-(a+b-c)}\\
   &\sim \Big(\Jcrit - \JU \Big)^{-(a+b-c)} \; .}
\end{equation}
Therefore, within the scope of hypergeometric analytic continuation by means of ${_2F_1}$, the divergence exponent of $c_{2k}$ at $\Jcrit$ is estimated by
\begin{equation}\label{eq:exponent_of_a_Gaussian_hypergeometric_function}
 \epsilon_{2k} = a + b - c \; .
\end{equation}
When employing the generalized hypergeometric functions ${_{q+1}F_q}$ as defined in equation~\eqref{eq:definition_of_the_generalized_hypergeometric_function}, this estimate becomes
\begin{equation}\label{eq:exponent_of_a_generalized_hypergeometric_function}
 \epsilon_{2k} = a_1 + \cdots + a_{q+1} - b_1 - \cdots - b_q \; .
\end{equation}
In order to ensure the validity of the transformation formula~\eqref{eq:linear_transformation_formula_for_hypergeometric_functions} for arguments $z~=~J/U\Big/(J/U)_{\rm c}~>~1$ we have to add to $\JU$ an arbitrarily small imaginary part. Moreover, beyond the radius of convergence of its series representation, the hypergeometric functions ${_{q+1}F_q}$ adopt complex values. If real-valued quantities are required, we may utilize the replacement
\begin{equation}
 {_{q+1}F_q}(z) \longrightarrow \lim_{\epsilon\rightarrow0} \Big({_{q+1}F_q}(z + i \epsilon) + {_{q+1}F_q}(z - i \epsilon)\Big)\Big/2
\end{equation}
beyond the phase transition. This construction resolves both issues. In this work, however, we will not study the correlation functions in the superfluid phase.
\section{Phase diagram}
\label{sec:4}
We now report the results we have obtained for the phase diagram by evaluating the radius of convergence of the one-particle correlation function $c_2$ according to the three different schemes. This phase diagram has previously been calculated in various ways~\cite{SandersEtAl15,FreericksMonien94,CapogrossoSansoneEtAl08,FreericksMonien96,SantosPelster09,FreericksEtAl09} and, therefore, allows us to assess the quality of the respective scheme. While the process-chain approach easily allows us to calculate the phase diagram for arbitrarily high filling factors, this proves difficult with other methods, so that we focus our comparison on the first Mott lobe ($g = 1$).
\subsection{Shanks transformation}
\label{subsection:Shanks_transformation}
\begin{figure}[tb]
 \begin{center}
  \includegraphics[width=0.8\textwidth]{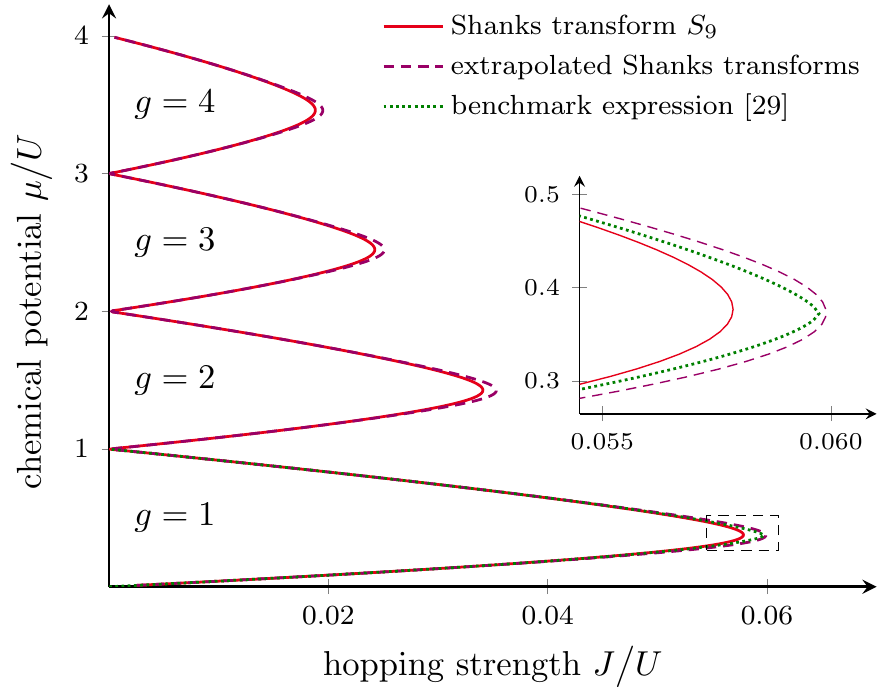}
 \end{center}
 \caption{(Color online) Zero-temperature phase diagram of the 2d Bose-Hubbard model as obtained from the Shanks transformation.}
 \label{fig:phase_diagram_comparison_with_Shanks_2d}
\end{figure}

When resorting to the Shanks transformation, we calculate the phase boundary by means of formula~\eqref{eq:Shanks_transform_phase_boundary} and extrapolate the results linearly in $1\big/\nu$, as indicated in Fig.~\ref{F_5}. We have depicted the highest numerically accessible approximation $S_9$ together with the extra\-polation in Fig.~\ref{fig:phase_diagram_comparison_with_Shanks_2d}.
Our results agree qualitatively very well with the benchmark expression for the phase boundary stated by Freericks~\etal~\cite{FreericksEtAl09}. Generally, the finite-order Shanks transform $S_9$ seems to significantly underestimate the boundary $\Jcrit$ around the tip of the Mott lobe, while the $1\big/\nu$-extrapolation seems to slightly overestimate the boundary away from the tip. Quantitatively, we find the tip of the first Mott lobe at $\muU = 0.3740$ compared to $\muU = 0.3724$ as stated by Freericks~\mbox{\etal,} and there is a small deviation between both estimates of the phase boundary with a maximum difference of $\Delta \Jcrit = 0.00105$, which corresponds to a relative difference of $\unit[1.9]{\%}$, half-way between $\muU = 0$ and the tip of the first Mott lobe.\\
From the trend in the finite orders available to our perturbational treatment we expect that the phase boundary derived from the Shanks transform $S_\nu$ by applying formula~\eqref{eq:Shanks_transform_phase_boundary} converges for $\nu \rightarrow \infty$ to a phase boundary very similar to the one obtained by Freericks~\etal. This is supported by the very good agreement emphasized in the inset of Fig.~\ref{fig:phase_diagram_comparison_with_Shanks_2d}.
\subsection{Pad\'e approximation}
\begin{figure}[tb]
 \begin{center}
  \includegraphics[width=0.8\textwidth]{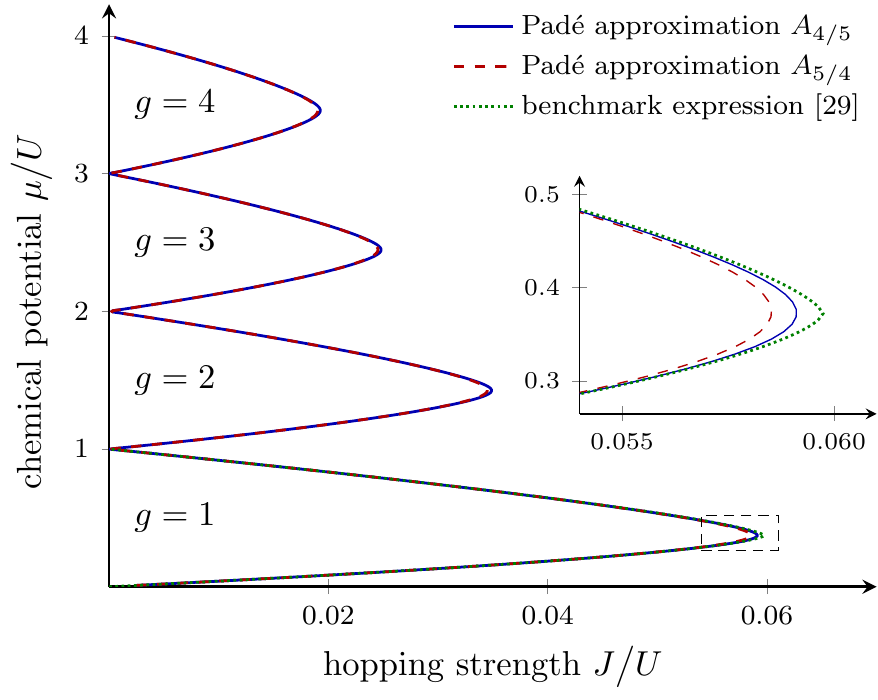}
 \end{center}
 \caption{(Color online) Zero-temperature phase diagram of the 2d Bose-Hubbard model as obtained from Pad\'e approximations $A_{L/M}$.}
 \label{fig:phase_diagram_comparison_with_Pade_2d}
\end{figure}

When employing the Pad\'e approximation, we obtain the phase boundary by calculating the zeros of the denominator $1 + \sum_{k=1}^M q_k \cdot (\JU)^k$ appearing in equation~\eqref{eq:definition_of_the_Pade_approximation}. As mentioned before, the smallest positive zero corresponds to the radius of convergence of $c_2$ determining the phase boundary. Depending on the choice of $L$ and $M$, which enumerate the number of degrees of freedom, we obtain approximations of varying quality. To gauge the best choice for $L$ and $M$ we have calculated the phase boundary at the tip of the lowest Mott lobe, and have arranged the results in Tab.~\ref{table:Pade_table_for_phase_boundary_from_c_2} in the form of a Pad\'e table.\\
\setlength{\tabcolsep}{3pt}
\renewcommand{\arraystretch}{1.25}
\begin{table}[ht]
 \begin{center}
 \begin{tabular}{x{30pt} x{20pt} | x{50pt} | x{50pt} | x{50pt} | x{50pt}}
                         & $M$   & 2       & 3       & 4      & 5 \\
  \multicolumn{2}{c|}{$L$}         &         &         &        & \\
  \hline
  \multicolumn{2}{c|}{2} & 0.04830 & 0.05706 & 0.05648 & 0.05697  \\
  \hline
  \multicolumn{2}{c|}{3} & 0.05609 & 0.05771 & 0.05789 & 0.05784  \\
  \hline
  \multicolumn{2}{c|}{4} & 0.05409 & 0.05788 & 0.05763 & 0.05911  \\
  \hline
  \multicolumn{2}{c|}{5} & 0.05715 & 0.05839 & 0.05823 & 0.05880
 \end{tabular}
 \end{center}
 \caption{Pad\'e approximations to $\Jcrit$ at the tip of the first Mott lobe, for various values of $L$ and $M$.}
 \label{table:Pade_table_for_phase_boundary_from_c_2}
\end{table}
Comparing these approximations with the value $\JcritQMC  = 0.05974(3)$ from QMC calculations~\cite{CapogrossoSansoneEtAl08}, we obtain the best results for $L = 4$ and $M = 5$. Apparently, from all Pad\'e approximations with an equal number of degrees of freedom $L + M + 1$, those with $M - L = 1$ yield the best results. Hence, we conjecture that Pad\'e approximations with $M - L = 1$ may have the correct asymptotics for $\JU \rightarrow \infty$.\\
We have displayed the phase boundary obtained by the Pad\'e approximations $A_{4/5}$ and $A_{5/4}$ in Fig.~\ref{fig:phase_diagram_comparison_with_Pade_2d}.
Both approximations agree very well with each other. In comparison with the benchmark we find a better quantitative agreement of $A_{4/5}$, as already anticipated from the Pad\'e table. Again, we find the tip of the first Mott lobe at $\muU = 0.3740$. The largest absolute difference between the phase boundary obtained from the Pad\'e approximation $A_{4/5}$ and the benchmark is found at the tip of the first Mott lobe, amounting to $\Delta \Jcrit = 0.00188$, which corresponds to a relative error of $\unit[3.2]{\%}$.\\
Again, examining the trend in the phase diagrams provided by finite-order Pad\'e approximations we expect convergence for $L + M \rightarrow \infty$ to a result very similar to the one obtained by Freericks~\etal~\cite{FreericksEtAl09}.
\subsection{Hypergeometric continuation}
\label{subsection:hypergeometric_function_continuation}
\begin{figure}[tb]
 \begin{center}
  \includegraphics[width=0.8\textwidth]{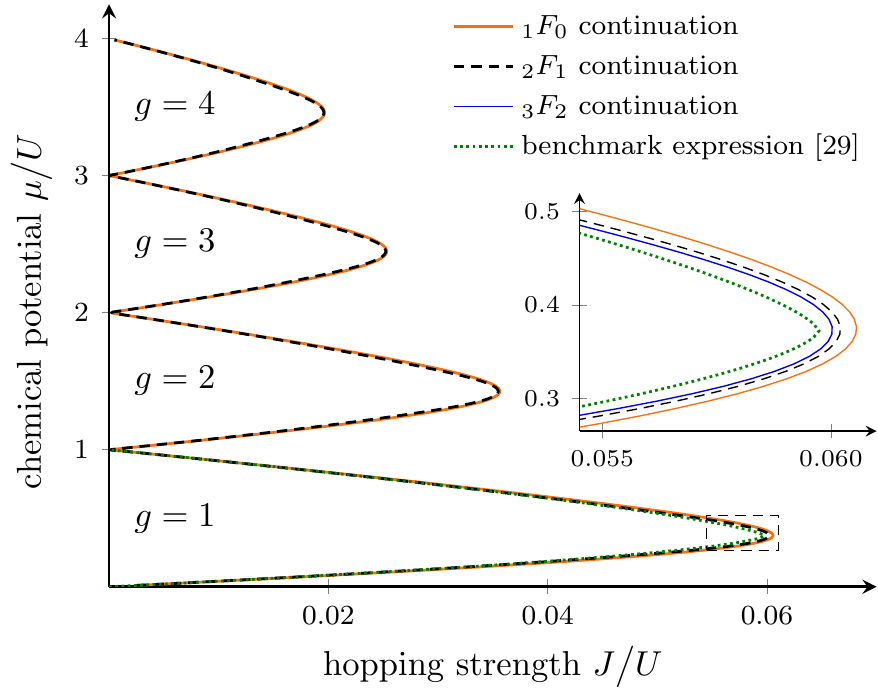}
 \end{center}
 \caption{(Color online) Zero-temperature phase diagram of the 2d Bose-Hubbard model as obtained from the hypergeometric functions ${_1F_0}$, ${_2F_1}$ and ${_3F_2}$.}
 \label{fig:phase_diagram_2d}
\end{figure}

\begin{figure}[tb]
 \begin{center}
  \includegraphics[width=0.8\textwidth]{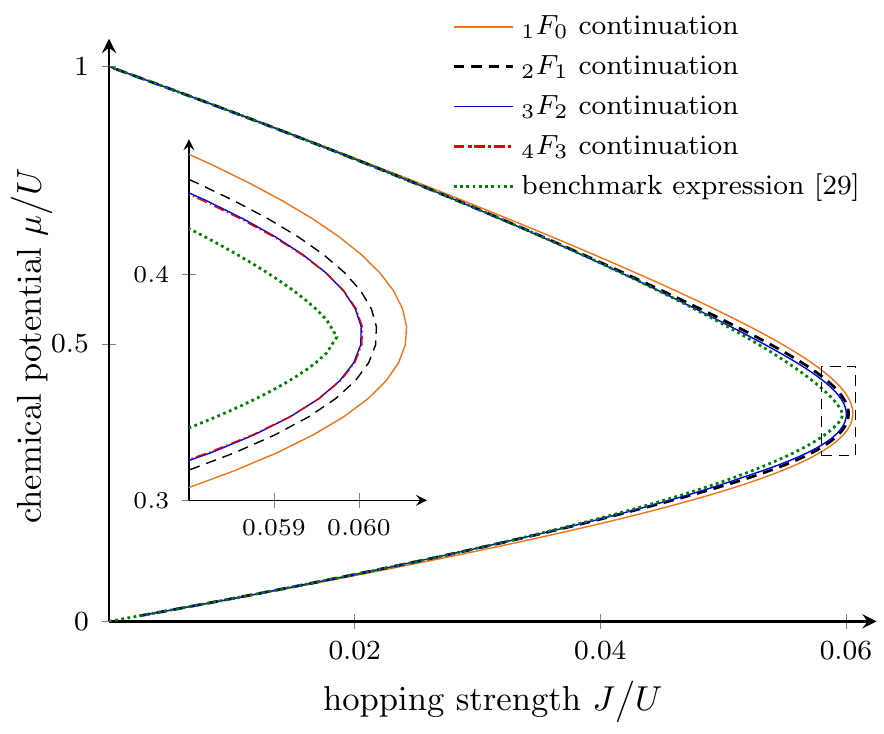}
 \end{center}
 \caption{(Color online) Zero-temperature phase diagram of the 2d Bose-Hubbard model as obtained from hypergeometric functions ${_{q+1}F_q}$ with $q=0,1,2,3$. Note that the boundaries for $q=2$ and $q=3$ almost coincide on the scale of this figure.}
 \label{fig:phase_diagram_2d_first_Mott_lobe}
\end{figure}

For each value of $\muU$ separately, we have fitted the ratios of subsequent coefficients $\alpha_2^{(\nu)}(\muU)\Big/\alpha_2^{(\nu-1)}(\muU)$ to the generalization of expression~\eqref{eq:ratio_in_the_ratio_fit} in order to obtain the parameters of the generalized hypergeometric functions ${_{q+1}F_q}$. The phase boundary $\Jcrit$ is then simply given by one of the fit parameters. In this manner we obtain the phase diagram depicted in Fig.~\ref{fig:phase_diagram_2d} for the first four Mott lobes, and magnified in Fig.~\ref{fig:phase_diagram_2d_first_Mott_lobe} for the first Mott lobe only.
Diagrams for ${_1F_0}$, ${_2F_1}$ and ${_3F_2}$ agree very well with each other, and with the benchmark expression. Exemplarily, we have calculated the largest difference between the phase boundary obtained by continuation based on ${_2F_1}$ and that benchmark. This gives $\Delta \Jcrit = 0.00151$, corresponding to a relative deviation of $\unit[2.8]{\%}$. When inspecting the different phase diagrams obtained with an increasing number of degrees of freedom using generalized hypergeometric functions ${_{q+1}F_q}$, the boundary $\Jcrit$ decreases and the difference between our findings and the benchmark becomes smaller. Similar to the Shanks transformation and the Pad\'e approximation the finite-order estimates seem to converge. In view of the minor differences between the phase diagrams obtained by ${_3F_2}$ and ${_4F_3}$, as shown in Fig.~\ref{fig:phase_diagram_2d_first_Mott_lobe}, it appears unlikely that the limit for $q \rightarrow \infty$ agrees exactly with the benchmark; however, the relative deviation is less than $\unit[2.0]{\%}$. Especially at the tip of the first Mott lobe the results hardly change when going from ${_3F_2}$ to ${_4F_3}$, with less than $\unit[0.25]{\%}$ relative deviation, and to ${_5F_4}$. The precise values of $\Jcrit$ at the tip provided by the various approximations are collected in Tab.~\ref{table:comparison_of_phase_boundary_at_tip_of_the_first_Mott_lobe}.
Thus, it appears that with ${_4F_3}$ we are very close to the limit $q \rightarrow \infty$, so that it is not necessary to calculate fits beyond ${_5F_4}$.
\subsection{Comparison}
\begin{figure}[tb]
 \begin{center}
  \includegraphics[width=0.8\textwidth]{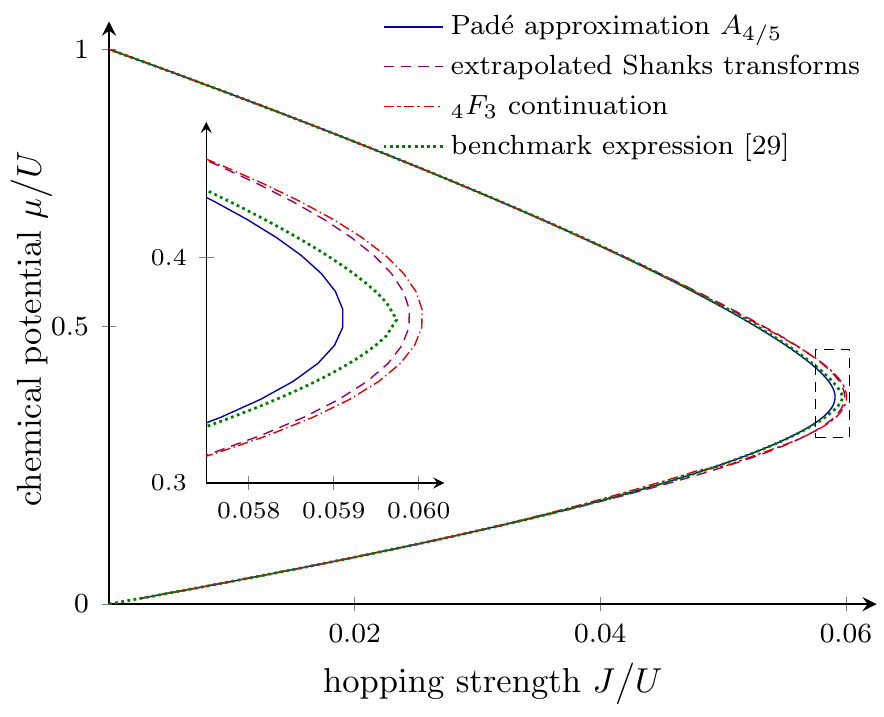}
 \end{center}
 \caption{(Color online) Zero-temperature phase diagram of the 2d Bose-Hubbard model as obtained from an extrapolation of Shanks transforms, the Pad\'e approximation $A_{4/5}$, and the hypergeometric function ${_4F_3}$.}
 \label{fig:phase_diagram_2d_first_Mott_lobe_comparison}
\end{figure}

Comparing the estimates for the phase boundary obtained from all three analytic continuation schemes, we conclude that they agree qualitatively very well with each other, yielding the well-known Mott lobes. As illustrated in Fig.~\ref{fig:phase_diagram_2d_first_Mott_lobe_comparison}, the finite-order Pad\'e approximation seems to underestimate the value $\Jcrit$ if compared to the other results or to the benchmark. A suitable extrapolation might improve the result and yield values closer to the ones obtained by the extrapolated Shanks transformation and hypergeometric function continuation. The latter ones agree quantitatively very well with each other, yielding values $\Jcrit$ slightly larger than the benchmark.
Tab.~\ref{table:comparison_of_phase_boundary_at_tip_of_the_first_Mott_lobe} gives a comparison between the phase boundary $\Jcrit$ at the tip of the first Mott lobe for some of our extrapolation schemes, and the QMC result $\JcritQMC = 0.05974(3)$.
\begin{table}[htb]
 \begin{center}
  \begin{tabular}{ c | c | c}
   method                                 & $\Jcrit$  & relative deviation\\[0.75ex] \hline
   Shanks transform $S_9$                 & \hspace*{5pt} $0.05786$ \hspace*{5pt} & $\unit[-3.15]{\%}$\\
   extrapolated Shanks transform          & \hspace*{5pt} $0.05989$ \hspace*{5pt} & $\unit[\phantom{-}0.25]{\%}$\\
   Pad\'e approximation $A_{4/5}$         & $0.05784$ & $\unit[         - 3.18]{\%}$\\
   ${_1F_0}$ continuation                 & $0.06056$ & $\unit[\phantom{-}1.37]{\%}$\\
   ${_2F_1}$ continuation                 & $0.06021$ & $\unit[\phantom{-}0.79]{\%}$\\
   ${_3F_2}$ continuation                 & $0.06003$ & $\unit[\phantom{-}0.49]{\%}$\\
   ${_4F_3}$ continuation                 & $0.06005$ & $\unit[\phantom{-}0.52]{\%}$\\
   ${_5F_4}$ continuation                 & $0.06004$ & $\unit[\phantom{-}0.65]{\%}$
  \end{tabular}
 \end{center}
 \caption{Comparison of the tip of the first Mott lobe, as resulting from various analytic continuation schemes, with the QMC result $\JcritQMC = 0.05974(3)$ from Ref.~\cite{CapogrossoSansoneEtAl08}.}
 \label{table:comparison_of_phase_boundary_at_tip_of_the_first_Mott_lobe}
\end{table}

The excellent agreement of the results derived from the novel hypergeometric function approach with those provided by the well-established Shanks transformation and Pad\'e approximation, and with the quantum Monte-Carlo simulation result, confirms the reliability of analytic continuation by means of hypergeometric functions for the computation of the phase boundary. This finding instills great confidence that hypergeometric analytic continuation also provides a reliable basis for the computation of divergence exponents.
\section{Divergence exponents of correlation functions}
\label{sec:5}
\begin{figure}[tb]
 \begin{center}
  \includegraphics[width=0.8\textwidth]{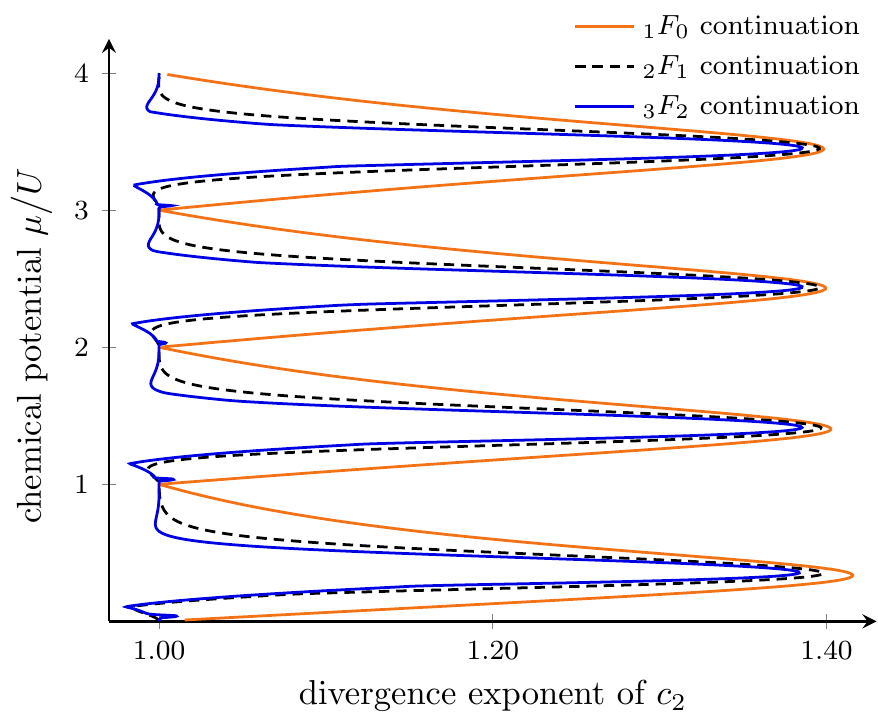}
 \end{center}
 \caption{(Color online) Divergence exponent $\epsilon_2$ of $c_2$ for the Mott insulator-to-superfluid transition, as obtained by hypergeometric continuation with ${_1F_0}$, ${_2F_1}$ and ${_3F_2}$, respectively.}
 \label{fig:exponent_diagram_first_four_Mott_lobes}
\end{figure}

\begin{figure}[tb]
 \begin{center}
  \includegraphics[width=0.8\textwidth]{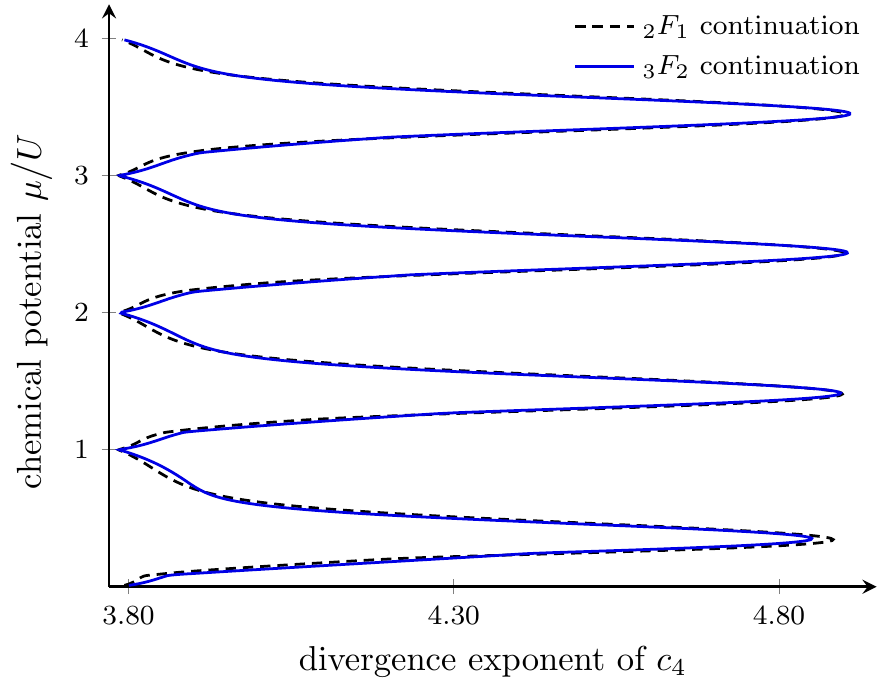}
 \end{center}
 \caption{(Color online) Divergence exponents $\epsilon_4$ of $c_4$ for the Mott insulator-to-superfluid transition, as obtained by hypergeometric continuation with ${_2F_1}$ and ${_3F_2}$, respectively. For ${_1F_0}$ we obtain qualitatively similar, but somewhat larger results.}
 \label{fig:exponent_for_c4_diagram_first_four_Mott_lobes}
\end{figure}

The key advantage of hypergeometric continuation, as compared to the Shanks or Pad\'e approach, lies in the fact that the hypergeometric functions themselves possess a non-trivial divergence exponent at the border of their convergence regime. Therefore, by fitting these functions to the perturbatively calculated coefficients $\alpha_{2k}^{(\nu)}$, the divergence exponents $\epsilon_{2k}$ of the correlation functions $c_{2k}$ are estimated directly from the parameters of the fit functions via equation~\eqref{eq:exponent_of_a_Gaussian_hypergeometric_function} for Gaussian hypergeometric functions, or equation~\eqref{eq:exponent_of_a_generalized_hypergeometric_function} for generalized hypergeometric functions. In this manner, the divergence exponents are obtained from the fits already presented in Sec.~\ref{subsection:hypergeometric_function_continuation}, without the need for additional work.\\
In Fig.~\ref{fig:exponent_diagram_first_four_Mott_lobes} we show the divergence exponent $\epsilon_2$ for scaled chemical potentials ranging from $\muU = 0$ to $\muU = 4$, as resulting from fits with ${_1F_0}$, ${_2F_1}$, and ${_3F_2}$.
Evidently, $\epsilon_2$ adopts its largest value at the tips of the Mott lobes, and the three estimates agree remarkably well there, whereas they differ notably in between.\\
Analogously, in Fig.~\ref{fig:exponent_for_c4_diagram_first_four_Mott_lobes} we display the divergence exponent $\epsilon_4$, as provided by ${_2F_1}$ and ${_3F_2}$. To obtain these fits each transition point $\Jcrit$ has been set to the value calculated before from $c_2$, instead of treating it as an independent fit parameter. This is justified by the knowledge that $c_2$ and $c_4$ share the same radius of convergence~\cite{SandersHolthaus17a}, but can also be checked numerically: If this assumption is not made, but $\Jcrit$ is determined independently from $c_2$ and $c_4$, the results agree to within less than $\unit[3]{\%}$.\\
To get an impression of how strongly our results depend on the parameter $q$ of the fitting functions ${_{q+1}F_q}$, Tab.~\ref{table:comparison_of_critical_exponent_at_tip_of_the_first_Mott_lobe} lists the divergence exponents at the tip of the first Mott lobe, as estimated for the $q$ numerically accessible to us. While the results for $\epsilon_2$ are quite stable, similar stability for $\epsilon_4$ is achieved only with $q = 1$ and $q = 2$, whereas the value for $q = 0$ appears to be somewhat off, indicating that ${_1F_0}$ does not offer a sufficient number of degrees of freedom.\\
\begin{table}[htb]
 \begin{center}
  \begin{tabular}{ c | c | c}
   method                                             & \hspace*{20pt}$\epsilon_2$\hspace*{20pt} & \hspace*{20pt}$\epsilon_4$\hspace*{20pt} \\[0.75ex]
   \hline
   \hspace*{5pt}${_1F_0}$ continuation\hspace*{5pt}   & $1.405$  & $5.515$ \\
   ${_2F_1}$ continuation                             & $1.390$  & $4.844$ \\
   ${_3F_2}$ continuation                             & $1.376$  & $4.762$ \\
   ${_4F_3}$ continuation                             & $1.383$  &         \\
   ${_5F_4}$ continuation                             & $1.391$  & 
  \end{tabular}
 \end{center}
 \caption{Comparison of the divergence exponents $\epsilon_2$ and $\epsilon_4$ at the tip of the first Mott lobe, as obtained by hypergeometric analytic continuation with various ${_{q+1}F_q}$.}
 \label{table:comparison_of_critical_exponent_at_tip_of_the_first_Mott_lobe}
\end{table}
From the connection of the divergence exponents to the critical exponents of the Mott insulator-to-superfluid transition~\cite{SandersHolthaus17a}, we know that both $\epsilon_2$ and $\epsilon_4$ each adopt a common value at all lobe tips. This expectation is confirmed by Tab.~\ref{table:comparison_of_critical_exponent_at_tips_of_different_Mott_lobe}, which has been compiled using ${_2F_1}$, on the sub-percent accuracy level.
\begin{table}[htb]
 \begin{center}
  \begin{tabular}{ c | c | c }
   Mott lobe $g$               & \hspace*{20pt}$\epsilon_2$\hspace*{20pt} & \hspace*{20pt}$\epsilon_4$\hspace*{20pt} \\[0.75ex]
   \hline
   1                           & $1.390$                    & $4.844$  \\
   2                           & $1.393$                    & $4.881$  \\
   3                           & $1.394$                    & $4.892$  \\
   4                           & $1.396$                    & $4.900$
  \end{tabular}
 \end{center}
 \caption{Comparison of the divergence exponents $\epsilon_2$ and $\epsilon_4$ at the tips of the first four Mott lobes, derived from the Gaussian hypergeometric function ${_2F_1}$.}
 \label{table:comparison_of_critical_exponent_at_tips_of_different_Mott_lobe}
\end{table}
\section{Conclusions}
\label{sec:6}
The present study serves both a technical and a conceptual purpose. On the technical level, we have employed the strong-coupling perturbation series for the correlation functions~\eqref{eq:series_expansion_of_c_2k} of the two-dimensional Bose-Hubbard model at zero temperature for demonstrating how analytic continuation of divergent perturbation series by means of hypergeometric functions~\cite{MeraEtAl15,PedersenEtAl16a,SandersEtAl15} is achieved in practice. To this end, we have evaluated the strong-coupling series numerically to high orders, and fitted their coefficients to generalized hypergeometric functions ${_{q+1}F_q}$, providing $2q+2$ degrees of freedom. This necessarily enforces a compromise between the quest for high flexibility, as increasing with $q$, and the task to provide a correspondingly large number of data points, i.e., to compute high-order coefficients of the perturbation series. Fortunately, it turns out that already the Gaussian hypergeometric function $_2F_1$ with its four degrees of freedom is suitable for most purposes. By comparing the phase diagram of the Mott insulator-to-superfluid transition computed in this manner with the results provided by the well known Shanks and Pad\'e continuation techniques, we have confirmed that hypergeometric continuation constitutes an accurate and reliable tool.\\
On the conceptual level, it is of profound theoretical interest to explore just \emph{how} the strong-coupling perturbation series diverge at the phase boundary. Namely, the Mott insulator-to-superfluid transition shown by the two-dimensional Bose-Hubbard model generally is of the mean field type, with the exception of the tips of the Mott lobes, which constitute multicritical points~\cite{FisherEtAl89}. Thus, the question arises how the switch from ``mean field-like'' to ``multicritical'' in response to a variation of the chemical potential affects the perturbation series. Hypergeometric continuation is ideally suited to address this question in detail, since generalized  hypergeometric functions provide a tunable singularity which can be employed to determine the exponents which characterize the divergence of the correlation functions at the phase boundary. Our numerical results for the divergence exponents $\epsilon_2$ and $\epsilon_4$ of the one- and two-particle correlation function, respectively, furnish the starting point of a novel strategy for determining critical exponents~\cite{SandersHolthaus17a}.\\
Thus, the idea to utilize hypergeometric functions for analytic continuation of divergent perturbation series, originally put forward by Mera \etal~\cite{MeraEtAl15}, not only provides an efficient alternative to the tools already at hand, such as the Shanks and Pad\'e techniques, but it also provides insights not obtainable with these older methods. Similar to the present study, the solution of a quantum mechanical single-particle or many-body problem can often be separated into two different steps: (i) The evaluation of a system-specific perturbation series, and (ii) the analytic continuation of that series. While step~(i) may strongly depend on the physical nature of the respective system under study, hypergeometric continuation offers a powerful universal tool to perform step~(ii).
%
%
%
\ack
We would like to thank Christoph Heinisch for his valuable contributions to the numerical evaluation of the perturbation series. We also acknowledge CPU time granted to us on the HPC cluster HERO, located at the University of Oldenburg and funded by the DFG through its Major Research Instrumentation Programme (INST 184/108-1 FUGG), and by the Ministry of Science and Culture (MWK) of the Lower Saxony State.
\appendix
\section{Bose-Hubbard model in the limit $d\rightarrow\infty$}\label{appendix:analytic_calculations_in_low_order}
In this Appendix, we calculate exemplarily the first three coefficients $\alpha_2^{(\nu)}$ of the perturbation series for $c_2$, and utilize the Shanks transformation to obtain an estimate for the phase boundary. As detailed in Sec.~\ref{sec:2}, we have to evaluate the correlation function
\begin{equation}
 c_2(\muU,\JU) = \alpha_2^{(0)}\!(\muU) + \alpha_2^{(1)}\!(\muU) \, \JU + \alpha_2^{(2)}\!(\muU) \left(\JU\right)^2 + \cdots \;. \;
\end{equation}
The contribution of $\alpha_2^{(0)}(\muU)$, which corresponds to the energy shift proportional to $\eta^2$ and $\left(\JU\right)^0$, is interpreted to originate from a process chain creating a Bose particle at an arbitrary site $i$, and annihilating it at the same site. According to the Kato perturbation theory we have to account for both permutations: Starting from the ground state~\eqref{eq:MottS}, either a particle is first created and then a particle is annihilated. In this case the sum over all intermediate states $\ket{m}$ reduces to a single term, involving a state $\ket{\ldots,g,g+1,g,\ldots}$ with one additional particle at site $i$, giving
\begin{equation}
 \eqalign{
        &\bra{g} \annihilation{i} \; \sum_{m \ne g} \frac{\ket{m}  \bra{m}}{E_g - E_m}     \; \creation{i} \ket{g}\\
   = \; &\bra{g} \annihilation{i} \; \frac{\ket{\ldots,g,\overbrace{g+1}^{{\rm site}\;i},g,\ldots}\bra{\ldots,g,\overbrace{g+1}^{{\rm site}\;i},g,\ldots}}{E_g - E_{g+1}} \; \creation{i} \ket{g}\\
   = \; &\frac{g + 1}{\muU - g} \; .}
\end{equation}
Or a particle is annihilated first and then a particle is created,
\begin{equation}
 \eqalign{
        &\bra{g} \creation{i} \; \sum_{m \ne g} \frac{\ket{m}  \bra{m}}{E_g - E_m} \; \annihilation{i} \ket{g}\\
   = \; &\bra{g} \creation{i} \; \frac{\ket{\ldots,g,\overbrace{g-1}^{{\rm site}\;i},g,\ldots}\bra{\ldots,g,\overbrace{g-1}^{{\rm site}\;i},g,\ldots}}{E_g - E_{g-1}} \; \annihilation{i} \ket{g}\\
   = \; &\frac{g}{(g-1) - \muU} \; .} 
\end{equation}
Hence, the zeroth-order term in $(\JU)$ is given by the sum
\begin{equation}\label{eq:appendix_zeroth_order_term}
 \eqalign{
 \alpha_2^{(0)}(\muU)
  &= \frac{g}{(g-1) - \muU} + \frac{g + 1}{\muU - g}\\
  &= \frac{-(\muU + 1)}{\Big(\muU - (g-1)\Big) \cdot (g - \muU)} \; .}
\end{equation}
Analogously, we compute the first-order contribution: With $i$ and $j$ labeling neighboring sites, this requires the calculation of
\begin{equation}\label{eq:appendix_example_first_order_perturbation_term}
 \eqalign{
  \bra{g} \annihilation{j} \sum_{m \ne g} \frac{\ket{m}  \bra{m}  }{E_g - E_m} (-\JU \; \creation{j} \, \annihilation{i}) \sum_{m' \ne g} &\frac{\ket{m'}  \bra{m'}  }{E_g - E_{m'}} \; \creation{i} \ket{g}\\
   & = - \frac{(g + 1)^2}{(\muU - g)^2} \, \JU \; ,}
\end{equation}
together with the five other permutations of the processes. As there are $2d$ directions for the tunneling process from a fixed site $i$ to a neighboring site $j$ on a $d$-dimensional hypercubic lattice, we have to multiply their sum by $2d$, and obtain
\begin{equation}\label{eq:appendix_first_order_term}
 \alpha_2^{(1)}(\muU) = \frac{-2d \, (\muU + 1)^2}{\Big(\muU-(g-1)\Big)^2 \cdot (g-\muU)^2} \; .
\end{equation}
To second order in $\JU$ we have to consider the two diagrams depicted in Fig.~\ref{F_2}: Either the second tunneling process ends at the origin of the first one or it goes to a different lattice site. For both diagrams we have to count the number of corresponding paths on our $d$-dimensional hypercubic lattice: While the first diagram has a multiplicity of $2d$, the multiplicity for the second diagram is $2d \cdot (2d-1)$. For general spatial dimension $d$ both diagrams have to be taken into account, and we have done so in our numerical evaluation for the two-dimensional square lattice. Here, however, we focus on the limit $d\rightarrow\infty$, which allows us to neglect the first diagram. Evaluating and summing over all $4!=24$ permutations, we find
\begin{equation}\label{eq:appendix_second_order_term}
 \eqalign{
  \alpha_2^{(2)}(\muU)
   &=       \frac{- 2d (2d-1) \, (\muU + 1)^3}{\Big(\muU-(g-1)\Big)^3 \cdot (g-\muU)^3}\\
   &\approx \frac{-(2d)^2     \, (\muU + 1)^3}{\Big(\muU-(g-1)\Big)^3 \cdot (g-\muU)^3} \; .}
\end{equation}
With these first three coefficients~\eqref{eq:appendix_zeroth_order_term}, \eqref{eq:appendix_first_order_term} and \eqref{eq:appendix_second_order_term} we are now equipped to calculate an approximation for the phase boundary by means of the Shanks transformation. According to equation~\eqref{eq:Shanks_transform_phase_boundary} this approximation is given by
\begin{equation}
 \eqalign{
  \Jcrit
   &=       \frac{\alpha_2^{(1)}(\muU)}{\alpha_2^{(2)}(\muU)}\\
   &\approx \frac{\Big(\muU-(g-1)\Big) \cdot (g-\muU)}{2d \, (\muU + 1)} \; .}
\end{equation}
This result coincides with the well-known mean-field expression for the phase boundary~\cite{FisherEtAl89}. The surprising observation that the mean-field boundary, which is exact in the limit $d\rightarrow\infty$, can be extracted by applying the Shanks transformation from the first three coefficients~$\alpha_2^{(\nu)}$ only, can be understood from the previous work by Teichmann~\etal~\cite{TeichmannEtAl09}. These authors have shown that the perturbation series for $c_2(\muU,\JU)$ takes the form of a geometric series in the limit $d\rightarrow\infty$. Thus, the sequence has a geometrically vanishing transient and therefore is mapped to its proper limit by the Shanks transformation.
\section{Comparison between ratio fit and relative deviation fit}
\label{appendix:comparison_between_ratio_and_weighted}
\begin{figure}[tb]
 \centering
  \begin{minipage}{.45\linewidth}\centering
   \includegraphics[width=\textwidth]{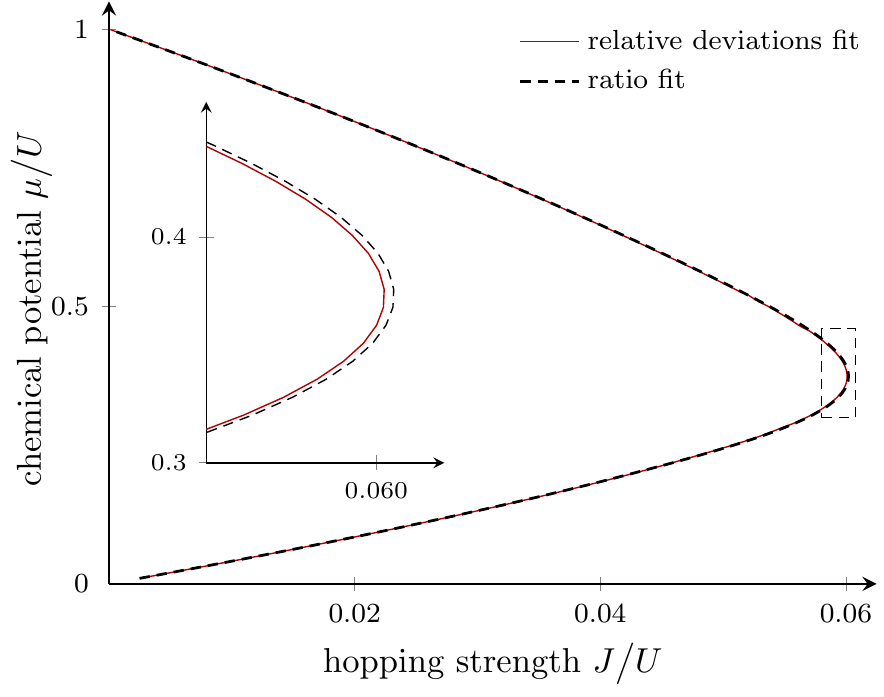}
  \end{minipage}
  \hfill
  \begin{minipage}{.45\linewidth}
   \includegraphics[width=\textwidth]{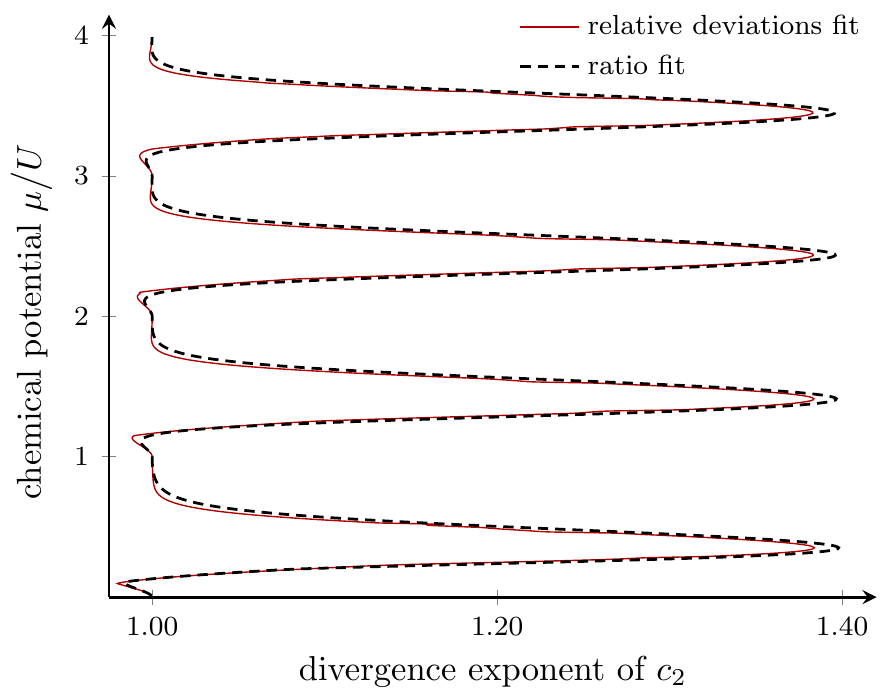}
  \end{minipage}
 \caption{(Color online) Zero-temperature phase diagram of the 2d Bose-Hubbard model (left) and divergence exponents $\epsilon_2$ for the Mott insulator-to-superfluid transition (right), as obtained from the hypergeometric functions ${_2F_1}$ with a least-squares fit minimizing the relative deviations, and with a least-squares fit to the ratios $\alpha_2^{(\nu)}\big/\alpha_2^{(\nu-1)}$.}
 \label{fig:phase_diagram_2d_comparison_between_weighted_and_ratio_fit}
\end{figure}

As described in subsection~\ref{subsection:hypergeometric_function_continuation} we have obtained the coefficients of the Gaussian hypergeometric functions ${_2F_1}$ by fitting the ratios~\eqref{eq:ratio_in_the_ratio_fit} to the calculated ratios $\alpha_2^{(\nu)}\big/\alpha_2^{(\nu-1)}$ of subsequent coefficients. However, we stress that we could have obtained very similar results if instead we had fitted the hypergeometric function coefficients directly to $\alpha_2^{(\nu)}$, weighting them in a suitable way. Without the introduction of weights most of the emphasis is put on the coefficients $\alpha_2^{(\numax{2}-1)}(\muU)$ and $\alpha_2^{(\numax{2})}(\muU)$ with the largest absolute values. As a consequence, the quality of the results decreases significantly: While the shape of the Mott lobes remains recognizable they lose their smoothness, and the divergence exponent appears to vary strongly for neighboring values of $\muU$.\\
In this Appendix we compare the previously employed ratio fits for the correlation function $c_2$ with a least-squares fit which minimizes the relative deviations
\begin{equation}
 \left|\frac{\alpha_2^{(0)}\!(\muU) \cdot \frac{(a)_\nu\,(b)_\nu}{\nu!\,(c)_\nu}\left(\frac{J/U}{(J/U)_{\rm c}}\right)^\nu - \alpha_2^{(\nu)}\!(\muU)}{\alpha_2^{(\nu)}\!(\muU)}\right| \; \; {\rm for} \; \; \nu \in \{1,..,\numax{2}\}
\end{equation}
with respect to $a$, $b$, $c$ and $\Jcrit$, effectively introducing the weights $1/\alpha_2^{(\nu)}$. Due to the significantly more complicated form of the fit function this is computationally more demanding, and takes considerably more CPU time to evaluate. However, as seen in Fig.~\ref{fig:phase_diagram_2d_comparison_between_weighted_and_ratio_fit}, both the resulting phase boundaries and the divergence exponents are in excellent agreement for the two strategies.
For a quantitative comparison we consider the respective fits at the tip of the first Mott lobe, and collect in Tab.~\ref{table:comparison_of_ratio_and_relative_deviation_fit} the parameters of ${_2F_1}$ obtained in either way.\\
\begin{table}[ht]
 \begin{center}
 \begin{tabular}{l|x{50pt} x{50pt} x{50pt} x{90pt} x{50pt}}
   strategy                & $a$ & $b$ & $c$ & $\epsilon_2 = a + b - c$ & $\Jcrit$ \\
   \hline
   ratio fit               & $1.399$ & $-0.7704$ & $-0.7663$ & $1.392$ & $0.06021$\\
   relative deviations fit & $1.383$ & $-0.5871$ & $-0.5763$ & $1.372$ & $0.06004$
 \end{tabular}
 \end{center}
 \caption{Comparison of parameters of ${_2F_1}$ fits to $c_2$ at the tip of first Mott lobe.}
 \label{table:comparison_of_ratio_and_relative_deviation_fit}
\end{table}
From these data we deduce that the corresponding estimates $\Jcritapprox{\rm ratio} = 0.06021$ and $\Jcritapprox{\rm relative} = 0.06004$ coincide to less than $\unit[0.5]{\%}$, and the divergence exponents $\epsilon_{2,\rm ratio} = 1.392$ and $\epsilon_{2,\rm relative} = 1.372$ deviate from each other by less than $\unit[1.4]{\%}$. This remarkably good agreement achieved with the two different fitting strategies further increases our confidence in the viability of the hypergeometric function approach for analytic continuation: Regardless of the particular fitting strategy one obtains stable estimates both for the phase boundary and even the divergence exponent.
%
%
%

\end{document}